\documentclass{scrartcl}
\usepackage{url}
\usepackage{graphicx}
\usepackage{amssymb}
\usepackage{color}
\usepackage{multicol}

\begin{document}
\title{Example Data Sets and Collections for BeSpaceD Explained}

\author{
Keith Foster and Jan Olaf Blech
}

\date{RMIT University, Melbourne, Australia}

\maketitle
\begin{abstract}
In this report, we present example data sets and collections for the
BeSpaceD platform. BeSpaceD is a spatio-temporal modelling and
reasoning software framework.
We describe the content of a number of the data sets and how the data was obtained.
We also present the programming API in BeSpaceD used to store and
access these data sets so that future BeSpaceD users can utilise the
data collections in their own experiments with minimal effort and
expand the library of data collections for BeSpaceD.
\end{abstract}

%------------------------------------------------------------------------------- Introduction

\section{Introduction}
This report continues the work on our spatio-temporal modelling and
reasoning framework BeSpaceD by adding a number of data sets useful for experimentation.
The data sets cover a varied range of data domains including three dimensional scans, robotics and environmental sensors.
We hope this diversity of data proves useful for future research in
BeSpaceD and for other applications using BeSpaceD.

BeSpaceD is our framework for spatio-temporal modelling and
reasoning\cite{bespaced1,bespaced0,bespacedURL}.  BeSpaceD is implemented using Scala
and is
characterised by (i) a description or modelling language that is based on abstract
datatypes, and (ii) means to reason about descriptions / models formulated in the description
language. 
The central datastructure in BeSpaceD is called {\tt Invariant}, a
formula that is supposed to hold for a system. Nevertheless,
invariants typically contain conditional parts. The main constructors
are realised as Scala abstract datatypes and some basic constructors
are provided below:

{\small
\begin{verbatim}
case class OR (t1 : Invariant, t2 : Invariant) extends Invariant;
case class AND (t1 : Invariant, t2 : Invariant)  extends Invariant;
case class NOT (t : Invariant) extends Invariant;
case class IMPLIES (t1 : Invariant, t2 : Invariant)  extends Invariant;

case class BIGOR (t : List[Invariant]) extends Invariant;
case class BIGAND (t : List[Invariant]) extends Invariant;

case class ATOM() extends Invariant;
case class TRUE() extends ATOM;
case class FALSE() extends ATOM;
\end{verbatim}
}

In addition to assumption logic, BeSpaceD features special
constructs for time and space as well as a variety of operators.

The BeSpaceD data set project is one component of the BeSpaceD platform which is growing to include projects related to distributed data collection, robotics, industrial automation and three dimensional scanning.
This report describes the application programming interface that the BeSpaceD platform provides for accessing the experimental data collections in Section~\ref{sec:api}.
Three dimensional scanning is covered in Section~\ref{sec:kin}.
Factory automation monitoring is covered in Section~\ref{sec:festo}.
Train occupancy monitoring is covered in Section~\ref{sec:train}.
Weather data is covered in Section~\ref{sec:wet}.

%------------------------------------------------------------------------------- Data Set API

\section{Data Set API}
\label{sec:api}
There are two types of users of the API for the data collections that
are subject to this report in BeSpaceD:
\begin{itemize}
\item Data Consumers - People who want to reuse a pre-existing data set.
\item Data Producers - People who have captured data and want to archive it for later reuse.
\end{itemize}

\subsection{Data Consumer API}
For users who only want to experiment in BeSpaceD with
pre-existing data sets, we provide a simple API that minimises the
effort so they can concentrate on the unique core of their research or
development activities.
In BeSpaceD data is provided as a formula, an abstract datatype using
Scala case classes. The formulas are supposed to hold for a system,
thus are invariants. The type is a Scala class named {\tt
  Invariant}. In order to attain such an {\tt Invariant} object containing a data set simply call one of the methods below:

\begin{verbatim}
    import BeSpaceDData
    
    Robotics.Lego.Trains.experiment1()
    Robotics.FestoMiniFactory.station1()
    Weather.SmartSpace.Melbourne.Aug_27_2015()
    Scan.Kinect.bottle()
    Scan.Kinect.obstacles()
\end{verbatim}

These methods return a single data set represented as an {\tt Invariant} object. In most cases this will be a BIGAND of other invariants.
These methods are dependent on archive files existing on disk. These
archives are not included in the BeSpaceDData project. % so they may fail.
In the current version the data archives are in the "data" directory under the root of the BeSpaceD repository.
One must clone the data directory from the git repository\footnote{\url{https://bitbucket.org/bespaced/bespaced-2016}} as well as the BeSpaceD project directory to ensure these access methods succeed.
If for some reason these archives are corrupted or deleted, these methods will throw an {\tt UnknownDatasetException}.

As the data sets can be very large, memory management is a concern.
The easy access methods above will load the data set from disk every time and will not cache the data at all in order to conserve memory.
When the object is de-referenced, the Java Virtual Machine will free (garbage collect) the memory eventually.
As an alternative, one may use the following fields in the {\tt InMemory} object. This simplifies access and only loads data on demand, however, beware that these objects will never be garbage collected.

{\footnotesize
\begin{verbatim}
import BeSpaceDData
    
object InMemory
{
  lazy val legoTrains = Robotics.Lego.Trains.experiment1()  
  lazy val festoStation1Scenario1 = Robotics.Festo.MiniFactory.station1.scenario1()
  lazy val festoStation1CapsBlocking = Robotics.Festo.MiniFactory.station1.capsBlocking() 
  lazy val melbourneWeather = Weather.SmartSpace.Melbourne.Aug_27_2015()
  lazy val kinectBottle = Scan.Kinect.bottle()
  lazy val kinectObstacles = Scan.Kinect.obstacles()
}
\end{verbatim}
}

With the above data set APIs we will ensure backward compatibility in future versions of BeSpaceDData.
If one prefers to cache the data to avoid multiple disk accesses, this can be done explicitly using the following methods. Note that one must deal with the data set names directly which are subject to change in the future.

{\small
\begin{verbatim}
    def loadAndCache(name: String): Option[Invariant]
    def findInCache(name: String): Option[Invariant]
    def findInCacheOrLoad(name: String): Option[Invariant]
    def findInCacheOrLoadAndCache(name: String): Option[Invariant]
\end{verbatim}
}

To make good use of the BeSpaceD data sets the structure of the data must be understood.
Currently there are four different data collections each with different data structures.
As of June 2016 we provide four different data collections each with its own structure and each collection may have multiple data sets.
The data collections and data sets will be added to over time.
A detailed explanation of the structure and content of each of these
collections can be found the respective sections.

\subsection{Data Producer API}
For users who have collected data we encourage  to donate it to the BeSpaceD open source project so that others in the community can make use of it to further the platform.
To this end we are also providing easy ways to archive  BeSpaceD data sets.
One can then commit this to a git fork of the BeSpaceD repository and then make a pull request to have it added to the community's data collection.

In order to create an archive of BeSpaceD data one first needs to name
it. We recommend using a prefix that is unique the respective organisation to avoid naming conflicts.
For example, here are the names of some of the data sets provided by
the Australia-India Centre for Automation Software Engineering
(AICAUSE) at RMIT University:

{\small
\begin{verbatim}
        aicause.lego.trains.experiment1
        aicause.festo.station1.Scenario1.20mins
        aicause.festo.station1.small.2capsBlocking
        aicause.smartspace.melbourne.2015.aug.27
        aicause.kinect.scan.bottle
        aicause.kinect.scan.obstacles
\end{verbatim}
}

Here the prefix of {\tt aicause} is reserved for their use, avoiding naming conflicts with any other contributors.
In addition the new data set may be added to the collections organised with a data domain ontology.
We have started the library of data collections in this maiden release of the BeSpaceDData project.
The beginnings of the data domain ontology is defined in Scala objects as follows:

\begin{multicols}{2}
{\small
\begin{verbatim}
  object Robotics
  {
    object Lego
    {
      object Trains
      object MindStorms
    }
    
    object ABB
    
    object Festo
    {
      object MiniFactory
    }
  }
  
  object Weather
  {
    object SmartSpace
    {
      object Melbourne
    }
  }
  
  object Scan
  {
    object Kinect
    object LeapMotion
  }
\end{verbatim}
}
\end{multicols}

After one has a name for the new data set, one needs to create the archive.
This is achieved by adding BeSpaceDData as a dependency to  a BeSpaceD-based project.
Then one will have the following methods available to call at the opportune time:
\begin{verbatim}
    def save(data: Invariant, name: String)
    def saveAndCache(data: Invariant, name: String)
\end{verbatim}
One will be collecting data and have it as an {\tt Invariant}
object. After completing the data set it needs to be saved by passing in the name of the data set.

The archive files are stored in the parent folder of the current working directory in a folder named {\tt data}.
%\begin{verbatim}
%    ../data/
%\end{verbatim}
When a  program is run within Eclipse and save() is called the above directory path matches the data directory in the BeSpaceD repository.
If a program is run with some other working directory, one needs to modify the data directory appropriately and/or copy data files to the BeSpaceD data directory.
To modify the data directory location the following source code line
in 
\begin{verbatim}
BespaceDData/src/BeSpaceDData/package.scala 
\end{verbatim}
needs to be changed:
\begin{verbatim}
    // IMPORTANT: Change this to your own path where the data is stored.
    private
    val archivePath = "../data"
\end{verbatim}

%------------------------------------------------------------------------------- Kinect Scanning

\section{Kinect Scanning Collection}
\label{sec:kin}

We gathered three dimensional scanning data from research projects
running in the VXLab facility \cite{vxlab}
using a Microsoft Kinect to scan various items.

\subsection{Data description}
This data describes the three dimensional surface of the objects.
In the Kinect scan data there are three different metrics obtained:

\begin{itemize}
\item
{\tt Coordinates of a point. } These are X, Y and Z coordinates of the area.
X and Y represent cartesian coordinates of a pixel in range from the Kinect box's point of view,
The Z value is a depth measurement.
\item 
{\tt UV values of a point. } The U and V values represent the texture coordinate mappings for 
      each pixel in the depth frame.
\item 
{\tt The color of a point. } These are R, G, B values of the color frame data.
\end{itemize}

The Kinect produces floating point values, however, in our adapter we converted these values into integers using approximation.

\begin{figure*}
\centering
\includegraphics[width=.35\textwidth]{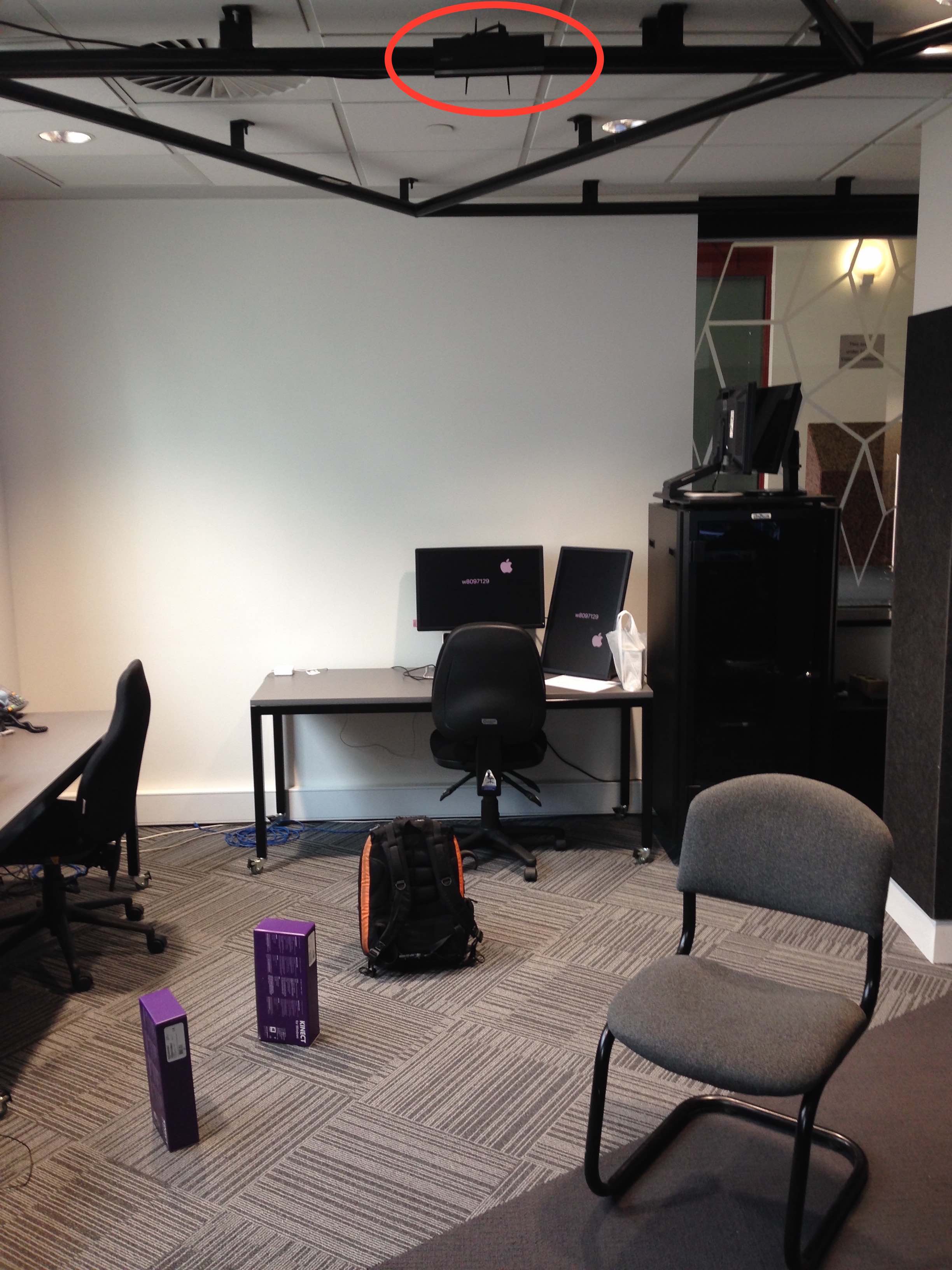}
\includegraphics[width=.62\textwidth]{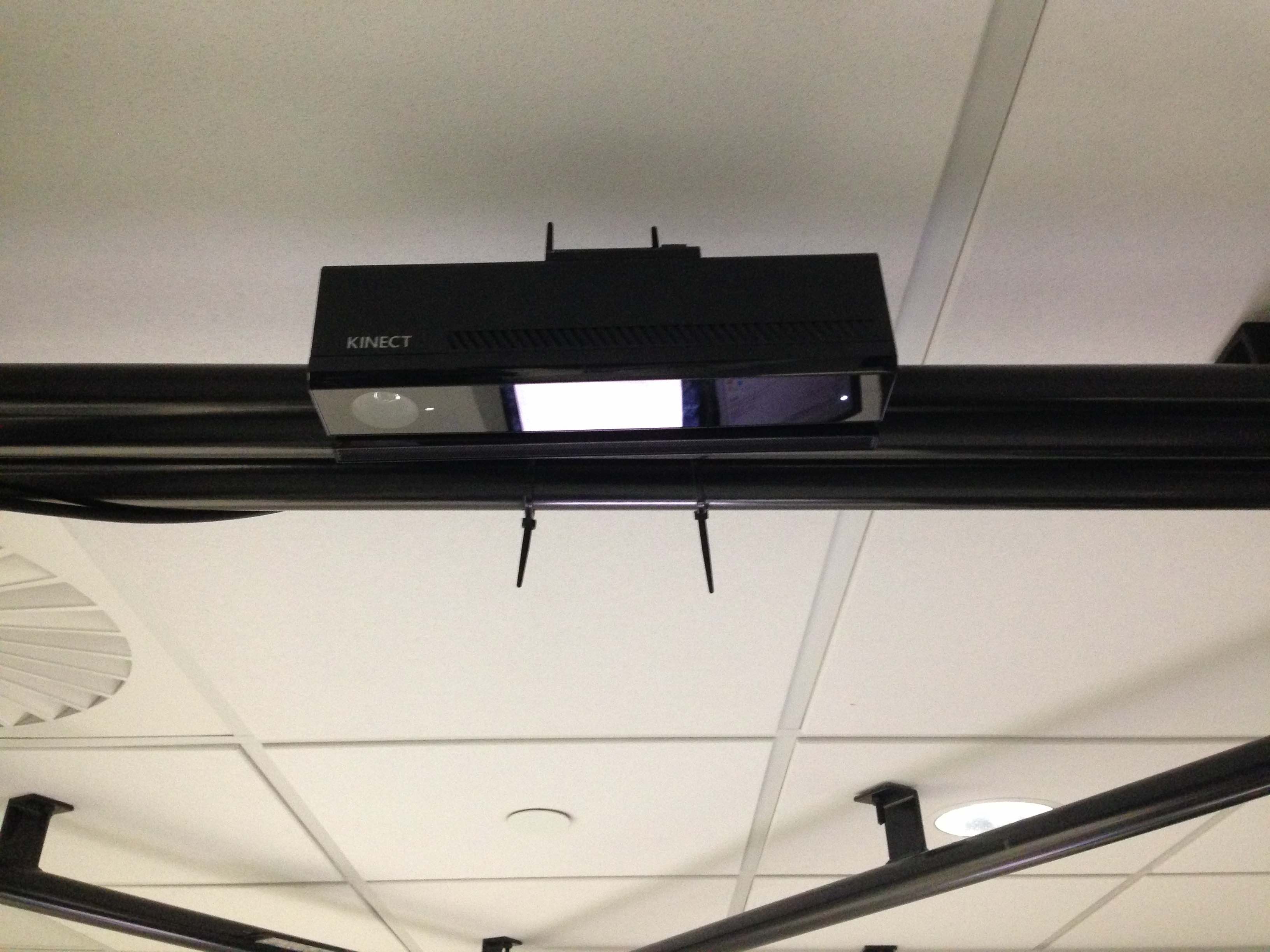}
\caption{Our Kinect Setup(left) and a closeup of the Kinect secured to the ceiling(right)}
\label{fig:kin}
\end{figure*}

In order to capture this data we positioned a Kinect device on the ceiling of our lab
facing down and positioned objects directly below it.
Figure~\ref{fig:kin} shows our Kinect setup.

\subsection{Data structure}
Each frame containing all points and colors is represented as a single IMPLIES invariant.
\begin{verbatim}
        IMPLIES(TimePoint(1429188806320), ...
\end{verbatim}
The time point is the premise and all the measurements for one frame are the conclusion.

All the measurements are represented as a BIGAND.
There are two kinds of measurements: points and colors, each represented as an implication.
The premise is an Owner Invariant describing the measurement type, e.g.
\begin{verbatim}
        Owner("Points")
        Owner("Colors")
\end{verbatim}
and the conclusion is all the measurements of that type contained in a BIGAND.
\begin{verbatim}
        IMPLIES(Owner("Points"), BIGAND(List(...
        IMPLIES(Owner("Colors"), BIGAND(List(...
\end{verbatim}

Each point measurement is represented as an IMPLES invariant.
Point measurements consist of two things that are associated : coordinates and UV values.
Implications are used to ensure the association between the co-ordinates and the UV values are retained.
The coordinate is represented as an Occupy3DPoint and the UV values are represented as a ComponentState storing a tuple of two integers.
\begin{verbatim}
        IMPLIES(Occupy3DPoint(-1,1,2),ComponentState(0,0))
\end{verbatim}

Each color measurement is represented as a ComponentState invariants.
Each component state value composes the RGB color values in a tuple of three integer values.
\begin{verbatim}
        ComponentState((41,49,39))
\end{verbatim}
Bringing all this together results in the following data structure
exemplified below:

{\small
\begin{verbatim}
IMPLIES(TimePoint(1429188806320),
BIGAND(List(
    IMPLIES(Owner("Points"), BIGAND(List(
       IMPLIES(Occupy3DPoint(-1,1,2),ComponentState(0,0)),
       IMPLIES(Occupy3DPoint(-1,1,1),ComponentState(0,0)),
       IMPLIES(Occupy3DPoint(-2,2,3),ComponentState(0,0)),
       IMPLIES(Occupy3DPoint(-1,1,2),ComponentState(0,0)),
       IMPLIES(Occupy3DPoint(-1,1,2),ComponentState(0,0)),
       IMPLIES(Occupy3DPoint(-1,0,1),ComponentState(0,0)),
       IMPLIES(Occupy3DPoint(-2,2,4),ComponentState(0,0)),
       IMPLIES(Occupy3DPoint(-1,1,2),ComponentState(0,0)),
       IMPLIES(Occupy3DPoint(-1,1,2),ComponentState(0,0)),
       IMPLIES(Occupy3DPoint(-1,1,2),ComponentState(0,0))
       ...
       )))

    IMPLIES(Owner("Colors"), BIGAND(List(
        ComponentState((41,49,39)), ComponentState((-1,38,46)),
        ComponentState((36,-1,38)), ComponentState((46,36,-1)), 
        ComponentState((38,46,36)), ComponentState((-1,38,46)),
        ComponentState((36,-1,40)), ComponentState((48,38,-1)),
        ComponentState((43,48,39)), ComponentState((-1,43,48))
        ...
        )))
)))
)
\end{verbatim}
}

\noindent This is just an excerpt of the first ten measurements of each metric of the actual data which contains a total of 217,088 3D points with corresponding UV coordinates and 2,764,800 color values.

\subsection{Data set access}
Currently we have sixteen Kinect scanning data sets two of which are
in the ontology. In order to access these data sets one can call any
of the following:

{\small
\begin{verbatim}
Scan.Kinect.bottle()
Scan.Kinect.obstacles()     // same as 1a below

loadOrThrow("aicause.kinect.scan.bottle")
loadOrThrow("aicause.kinect.scan.obstacles1")
loadOrThrow("aicause.kinect.scan.obstacles1a")
loadOrThrow("aicause.kinect.scan.obstacles2")
loadOrThrow("aicause.kinect.scan.obstacles3")
loadOrThrow("aicause.kinect.scan.obstacles4")
loadOrThrow("aicause.kinect.scan.obstacles5")
loadOrThrow("aicause.kinect.scan.obstacles6")
loadOrThrow("aicause.kinect.scan.obstacles7")
loadOrThrow("aicause.kinect.scan.obstacles8")
loadOrThrow("aicause.kinect.scan.obstacles9")
loadOrThrow("aicause.kinect.scan.obstacles10")
loadOrThrow("aicause.kinect.scan.obstacles11")
loadOrThrow("aicause.kinect.scan.obstacles12")
loadOrThrow("aicause.kinect.scan.obstacles13")
loadOrThrow("aicause.kinect.scan.obstacles14")
loadOrThrow("aicause.kinect.scan.obstacles15")
\end{verbatim}
}

\begin{figure}
\centering \includegraphics[width=.8\textwidth]{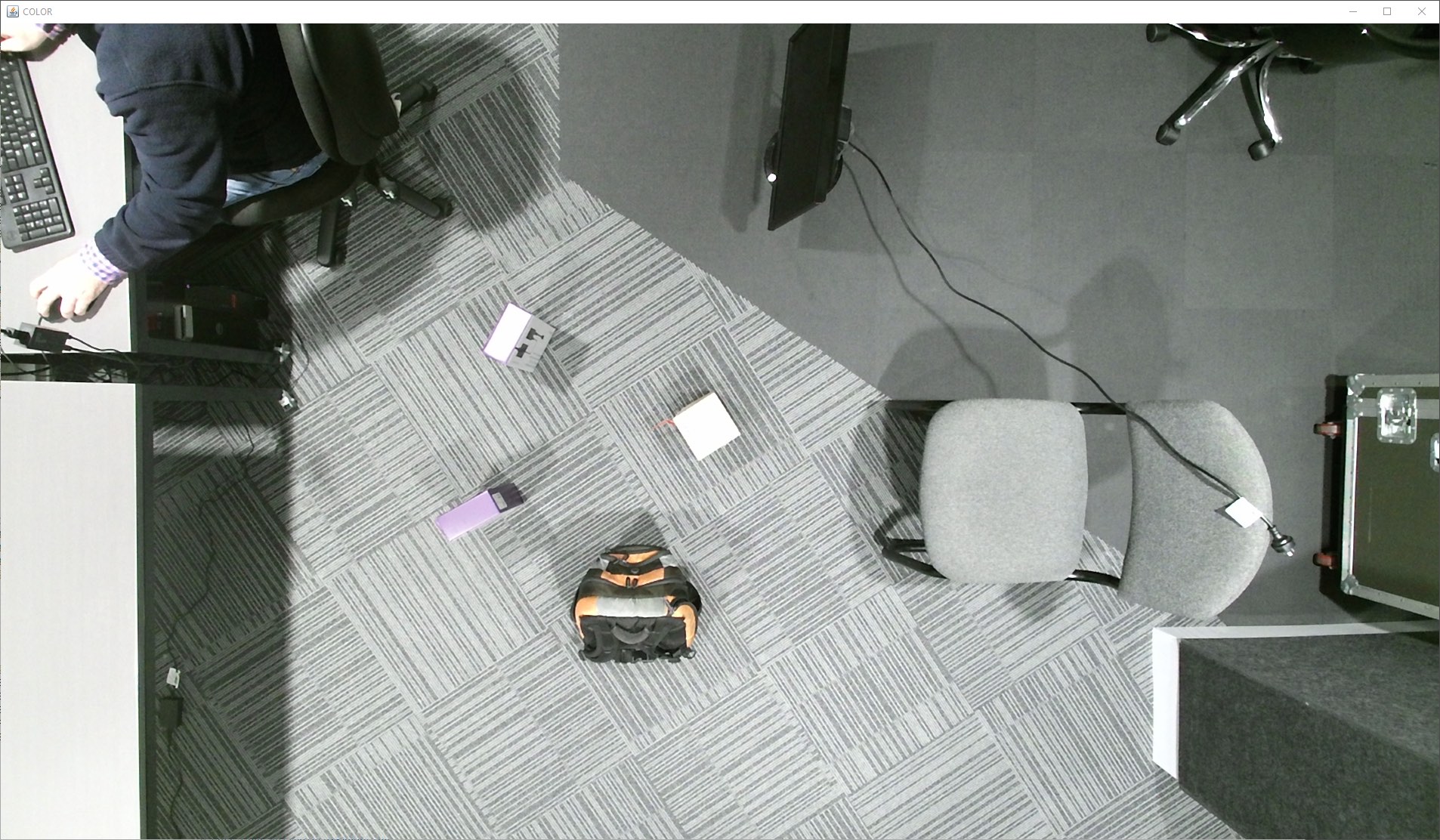}
\caption{Color data image for the obstacles1a data set}
\label{fig:ob1ac}
\end{figure}

\begin{figure}
% Skeleton
\centering\includegraphics[width=.6\textwidth]{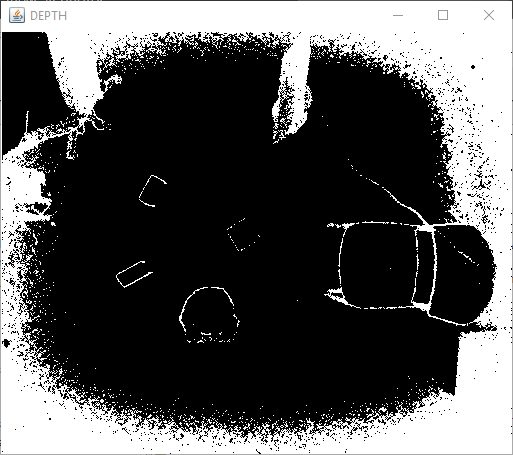}
\caption{Skeleton image for the obstacles1a data set}
\label{fig:ob1as}
\end{figure}

\begin{figure*}
\centering
\includegraphics[width=.3\textwidth]{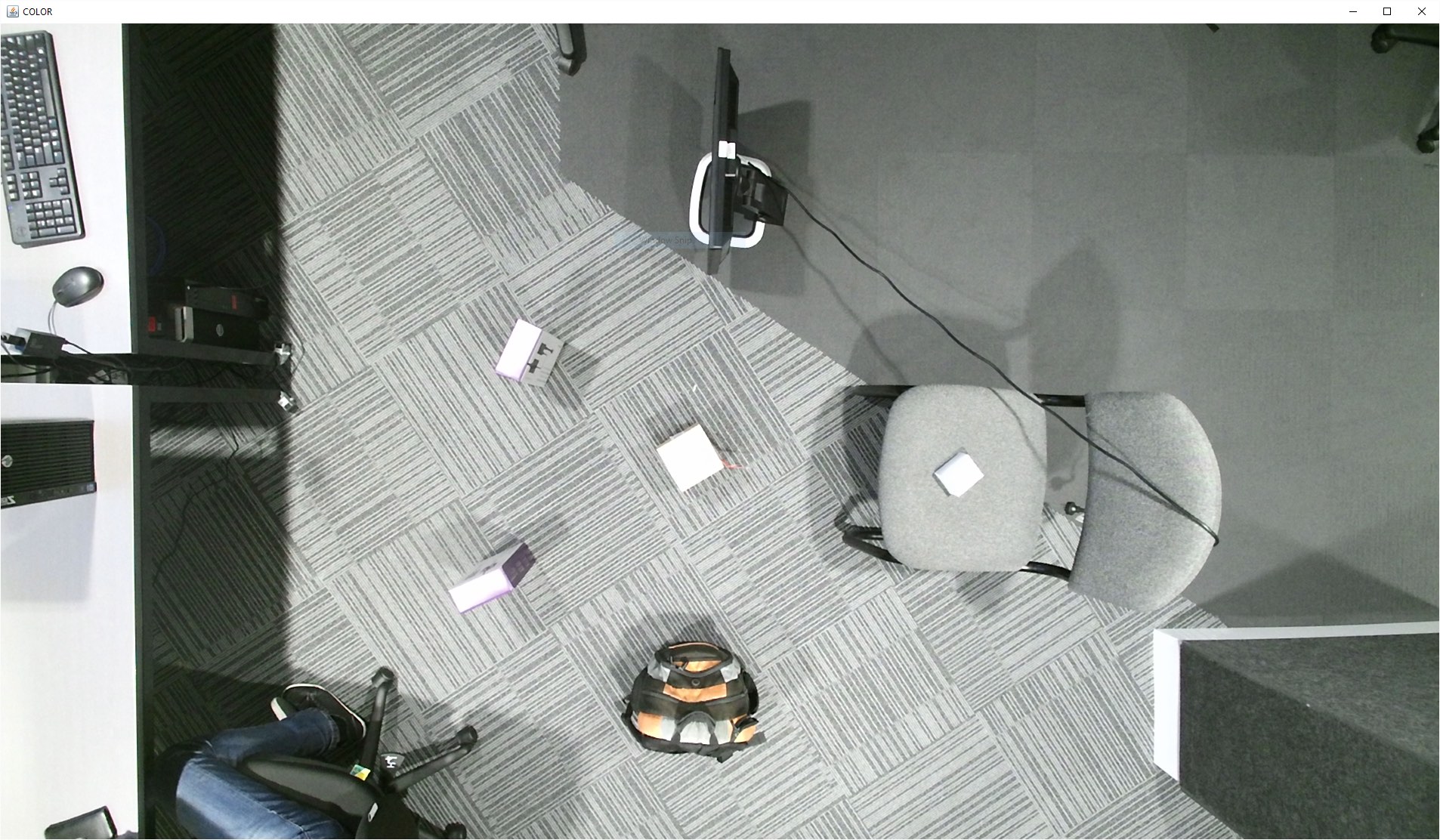}
\includegraphics[width=.3\textwidth]{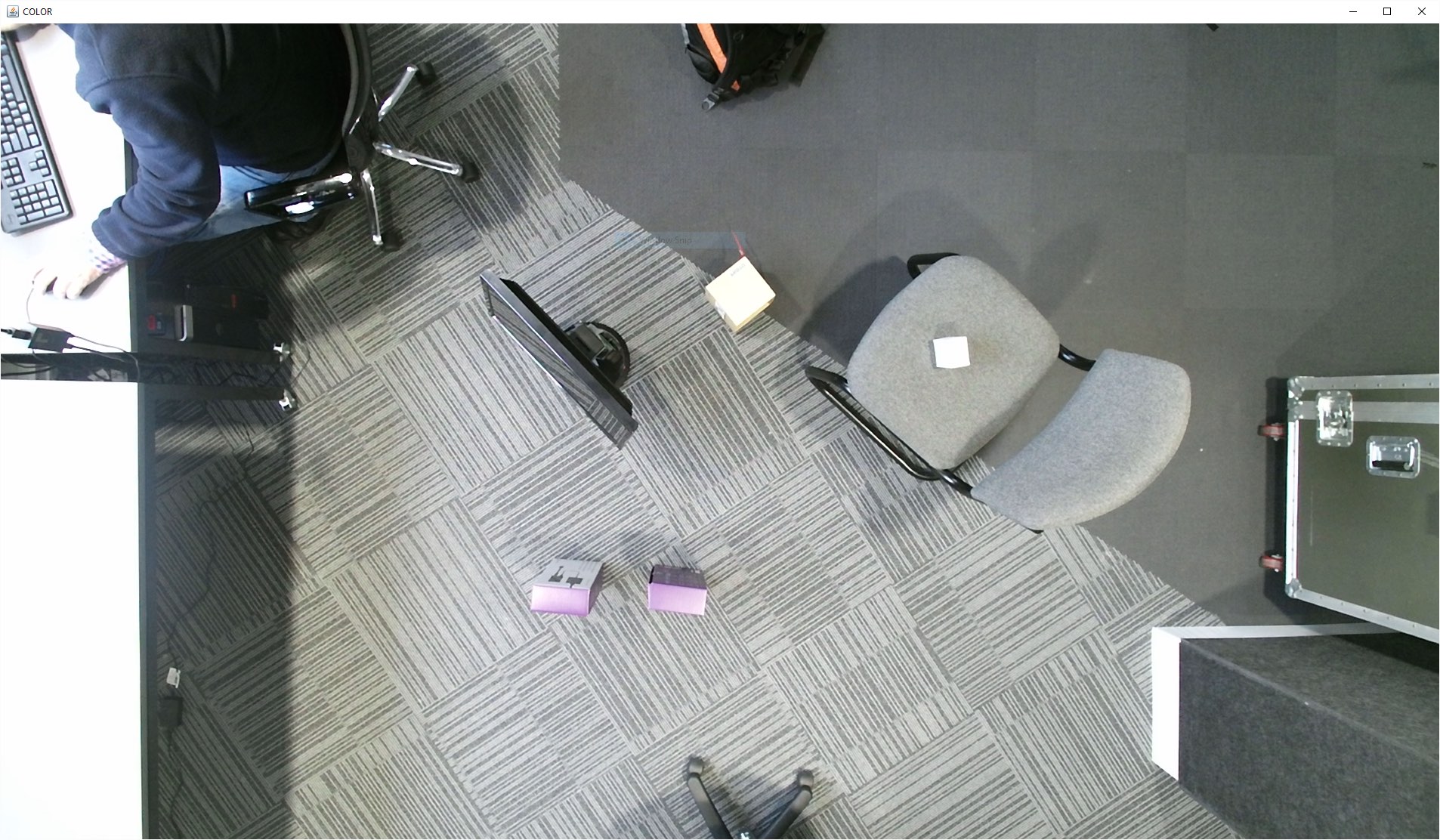}
\includegraphics[width=.3\textwidth]{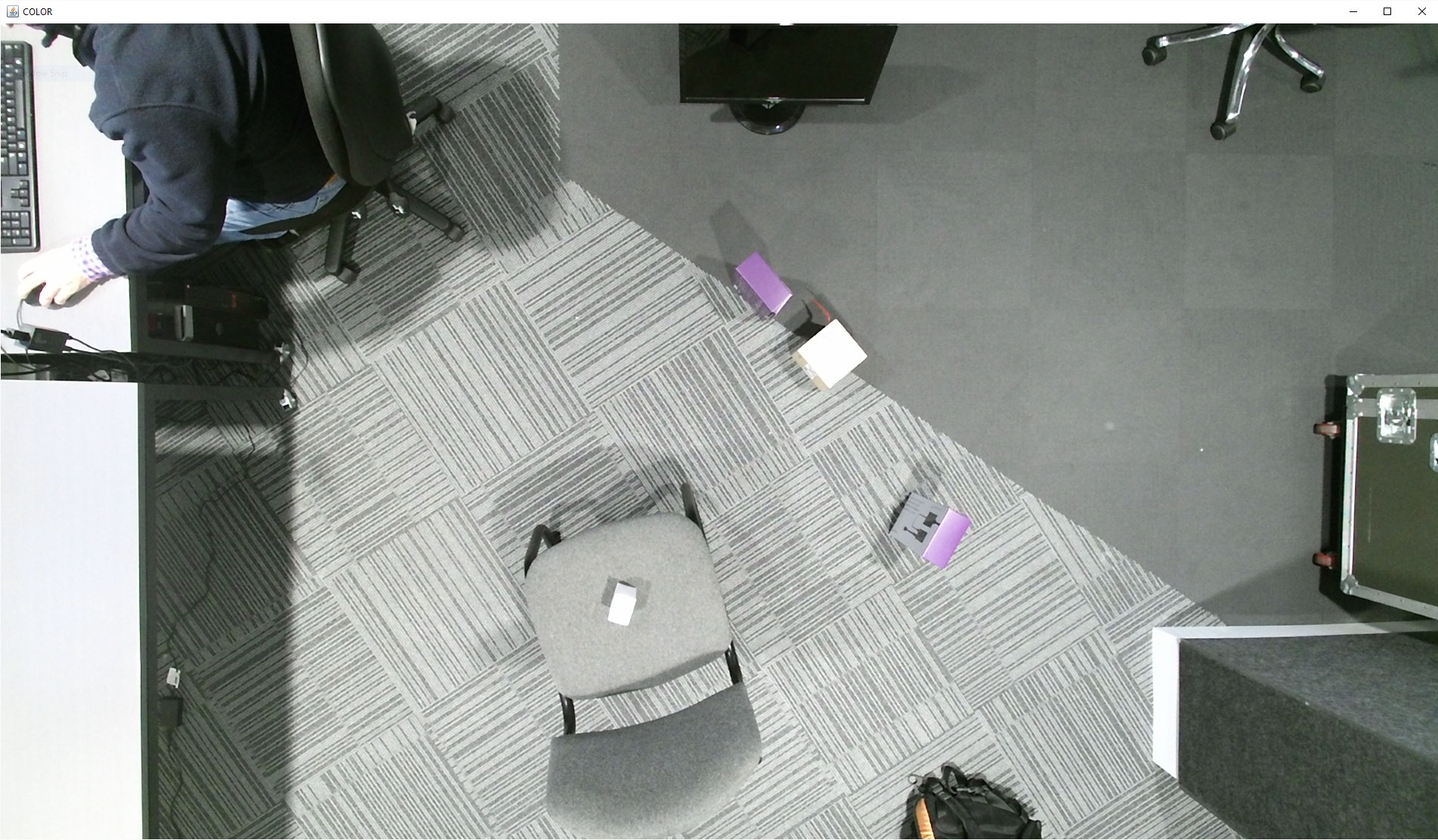}
\includegraphics[width=.3\textwidth]{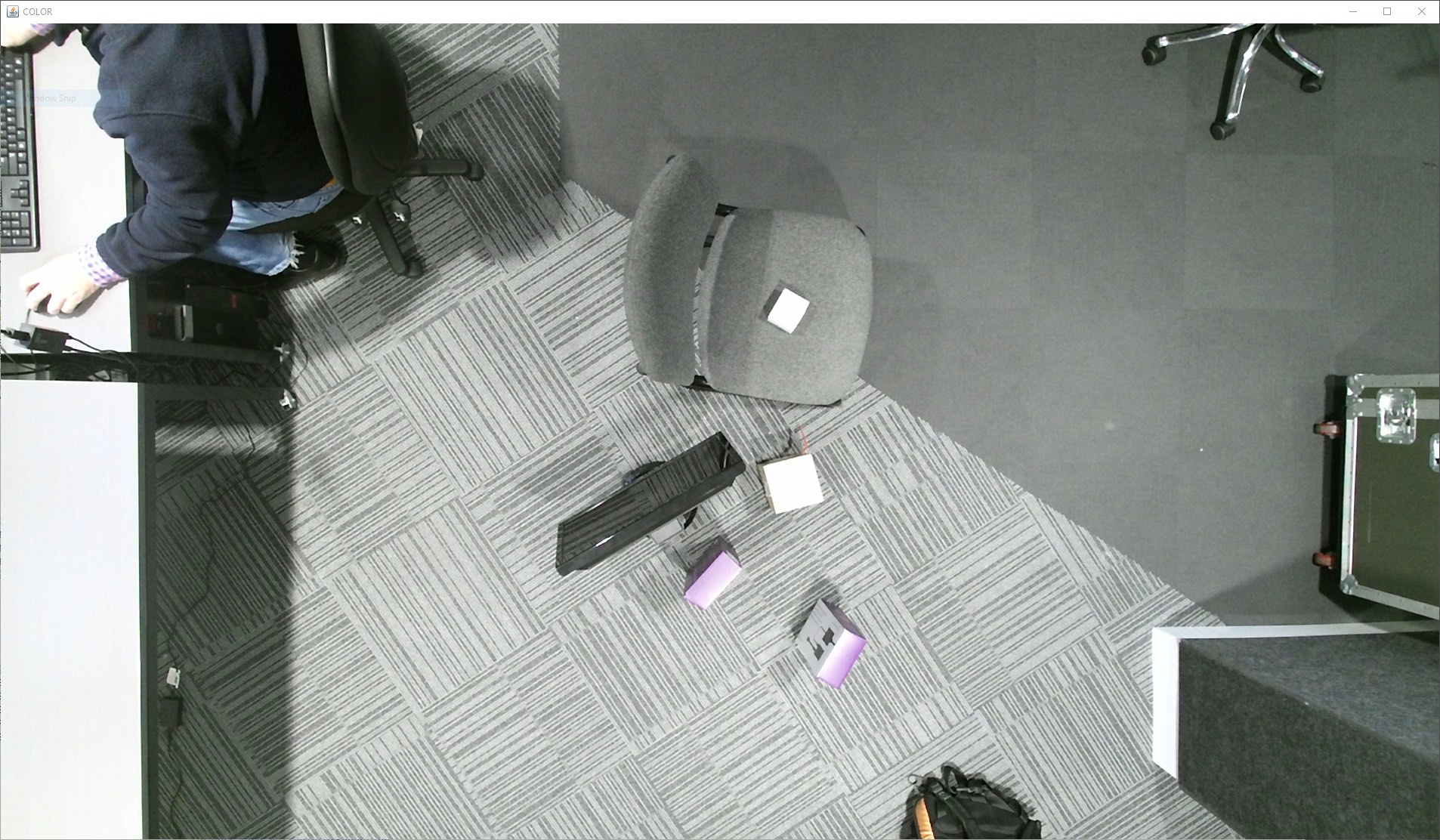}
\includegraphics[width=.3\textwidth]{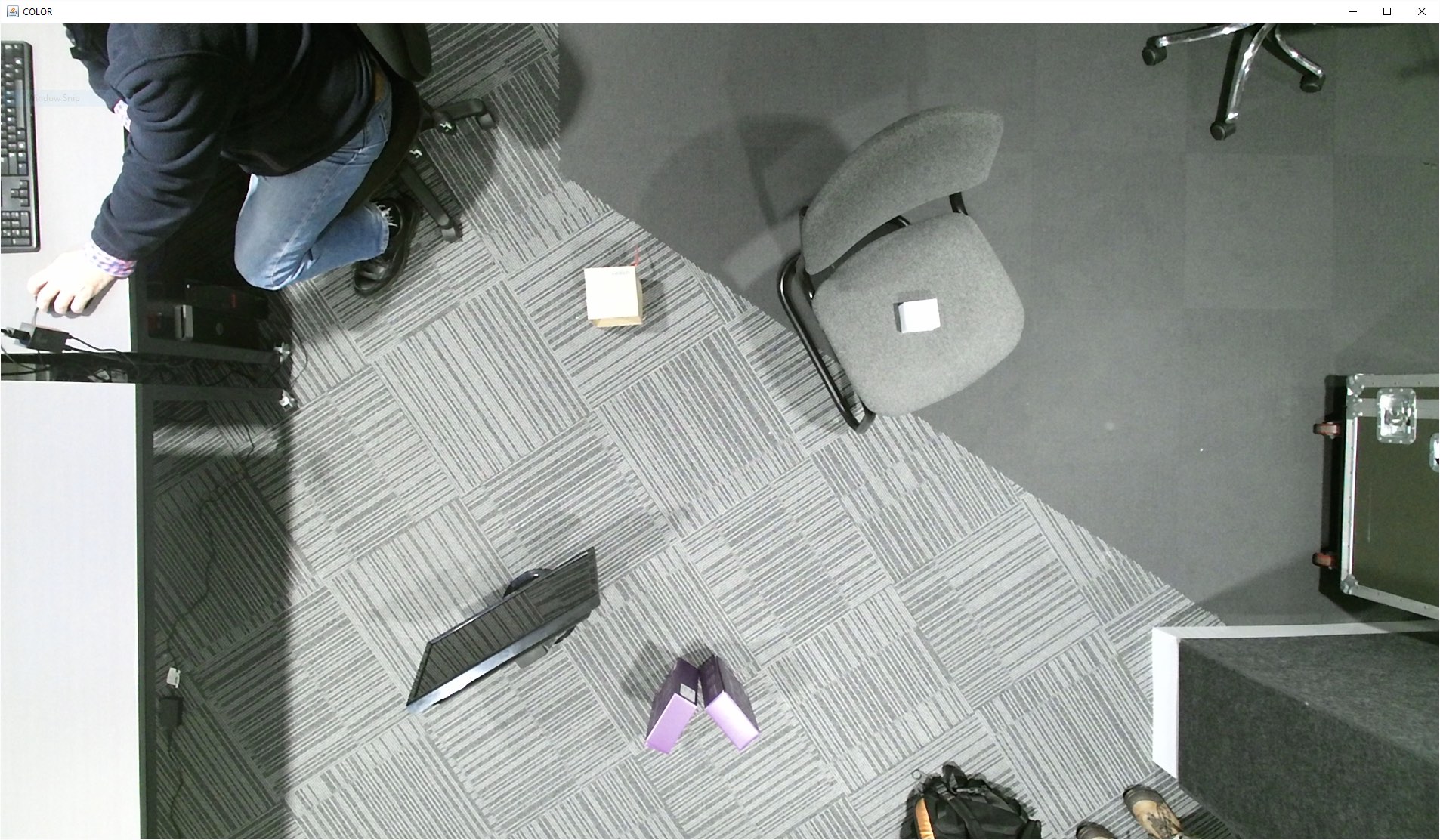}
\includegraphics[width=.3\textwidth]{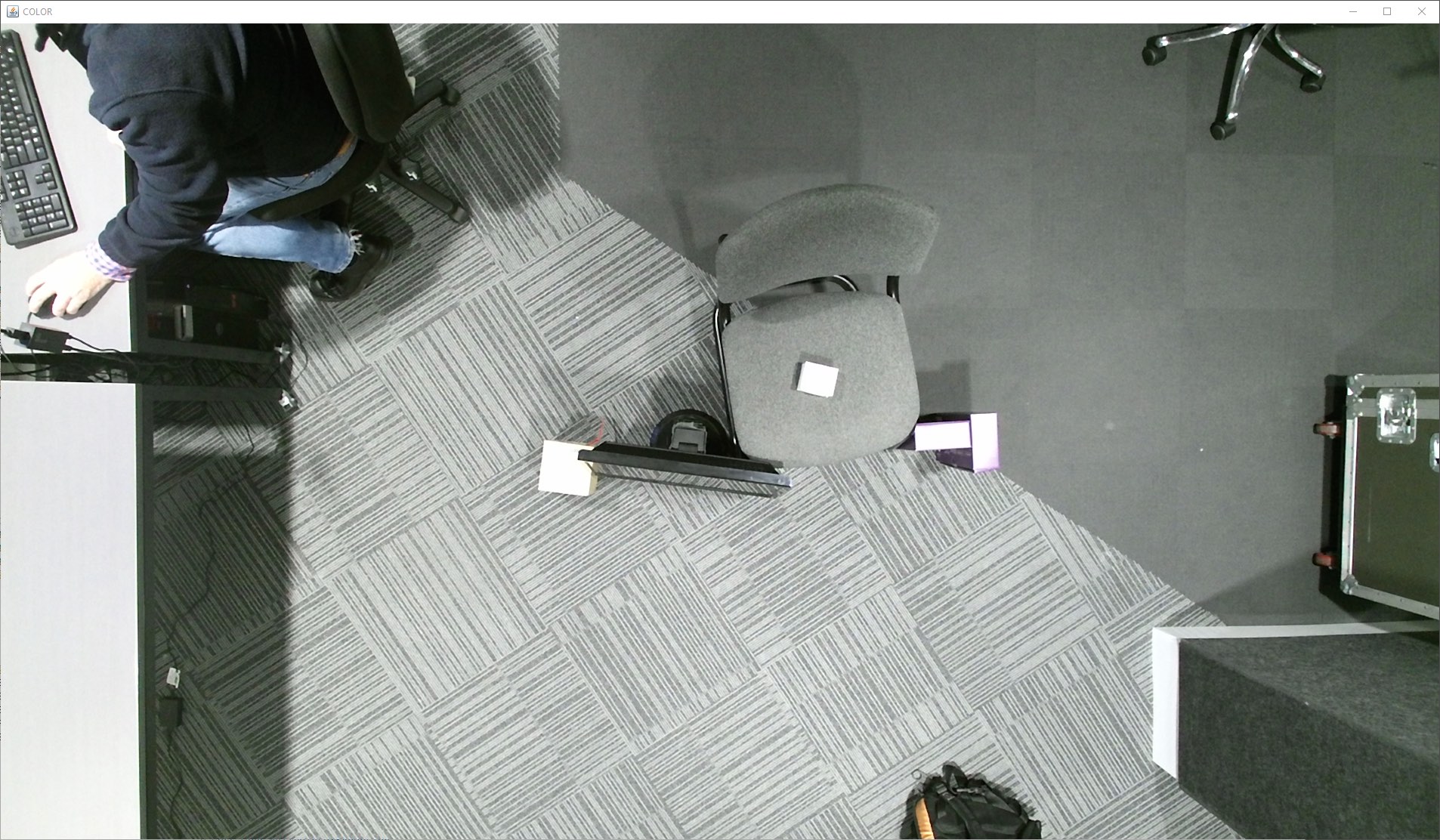}
\includegraphics[width=.3\textwidth]{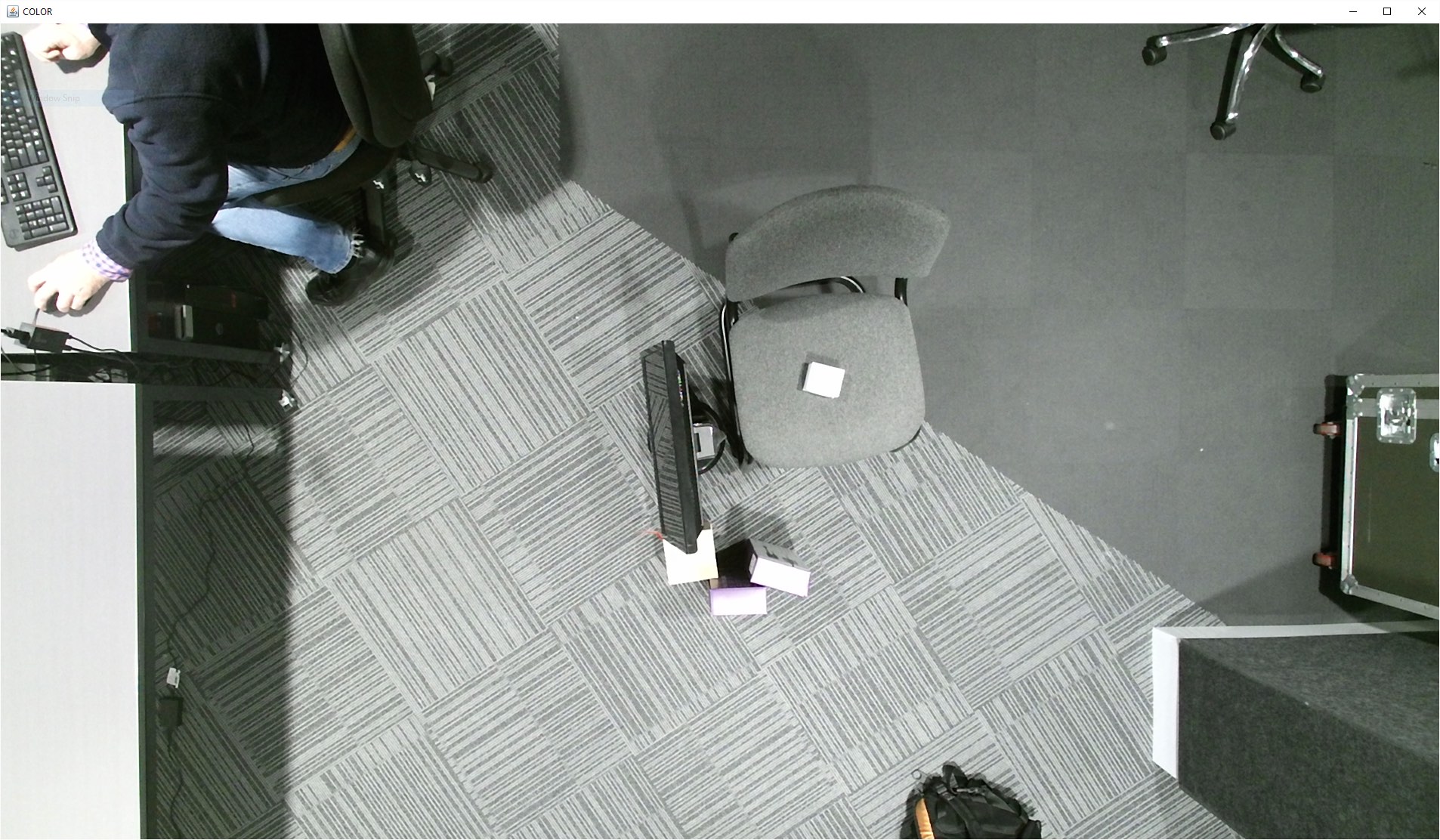}
\includegraphics[width=.3\textwidth]{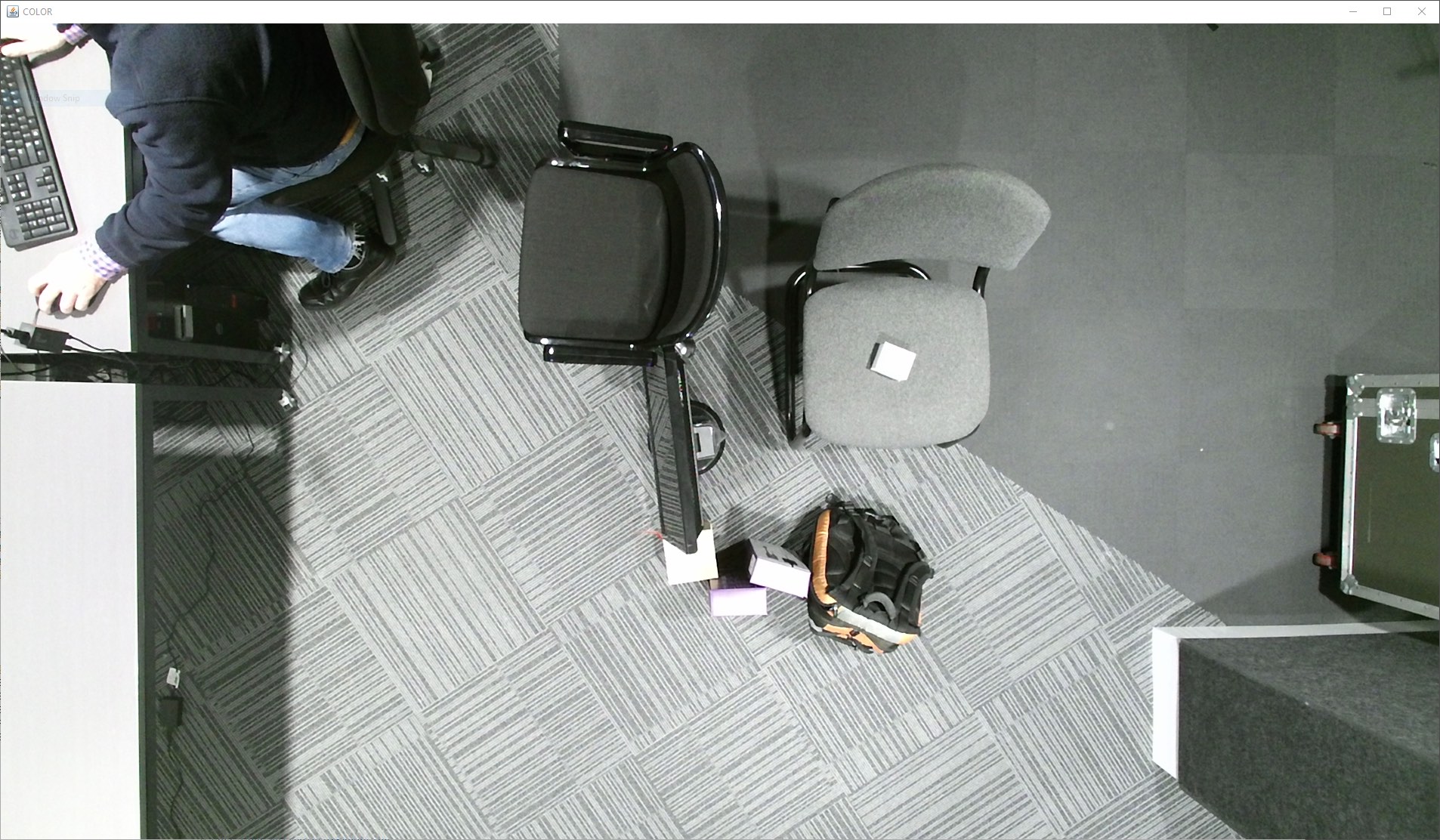}
\includegraphics[width=.3\textwidth]{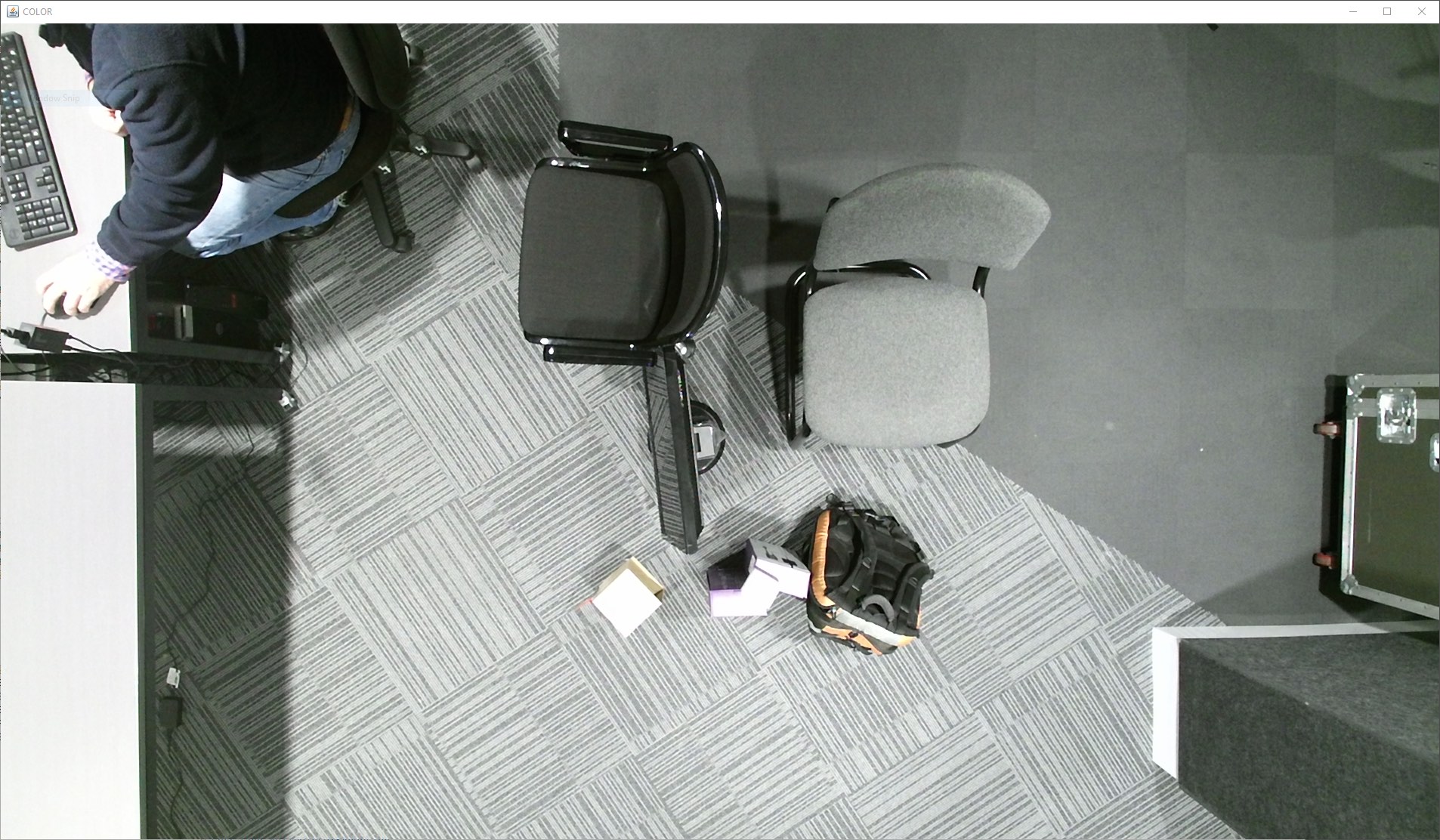}
\includegraphics[width=.3\textwidth]{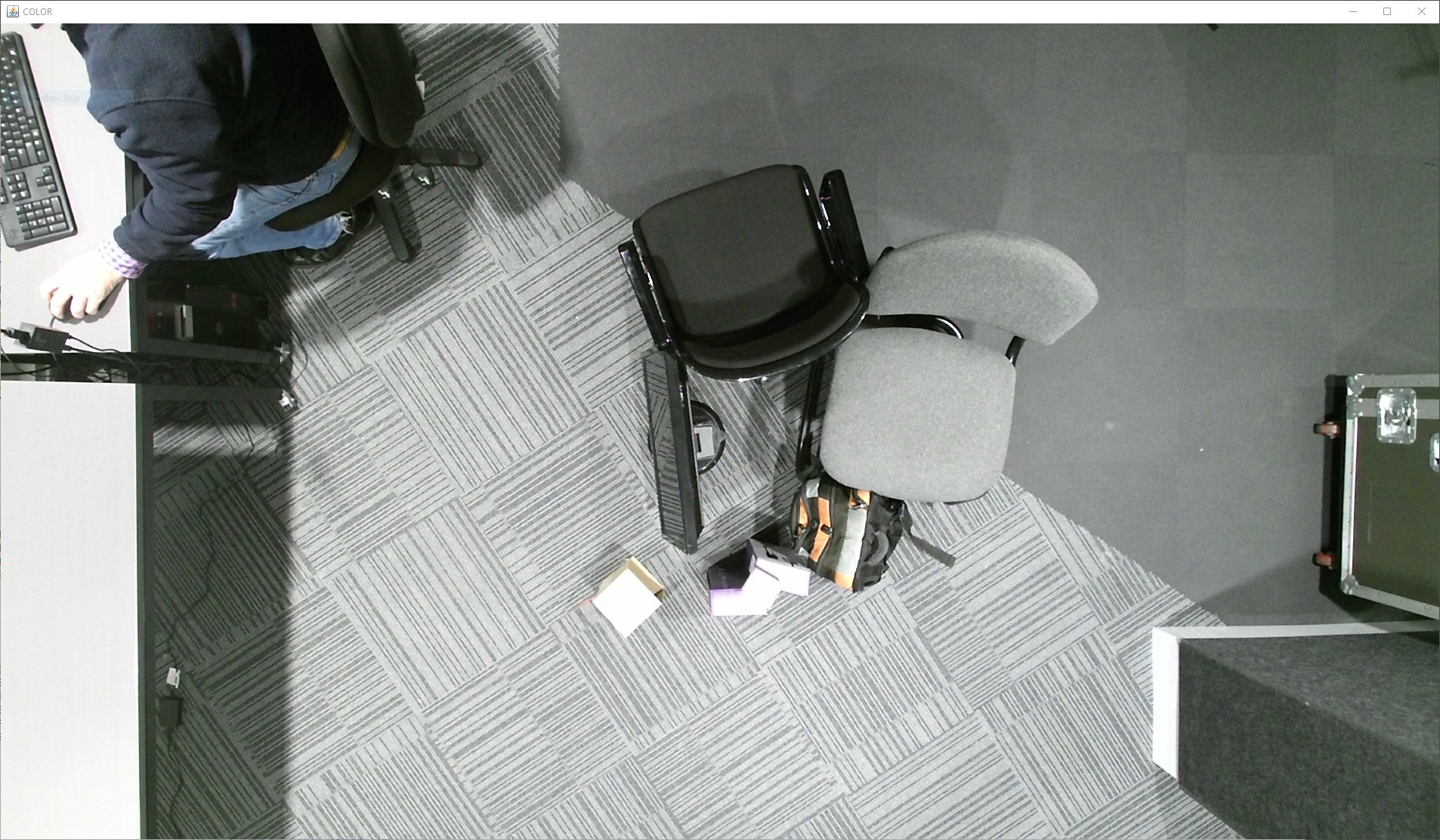}
\includegraphics[width=.3\textwidth]{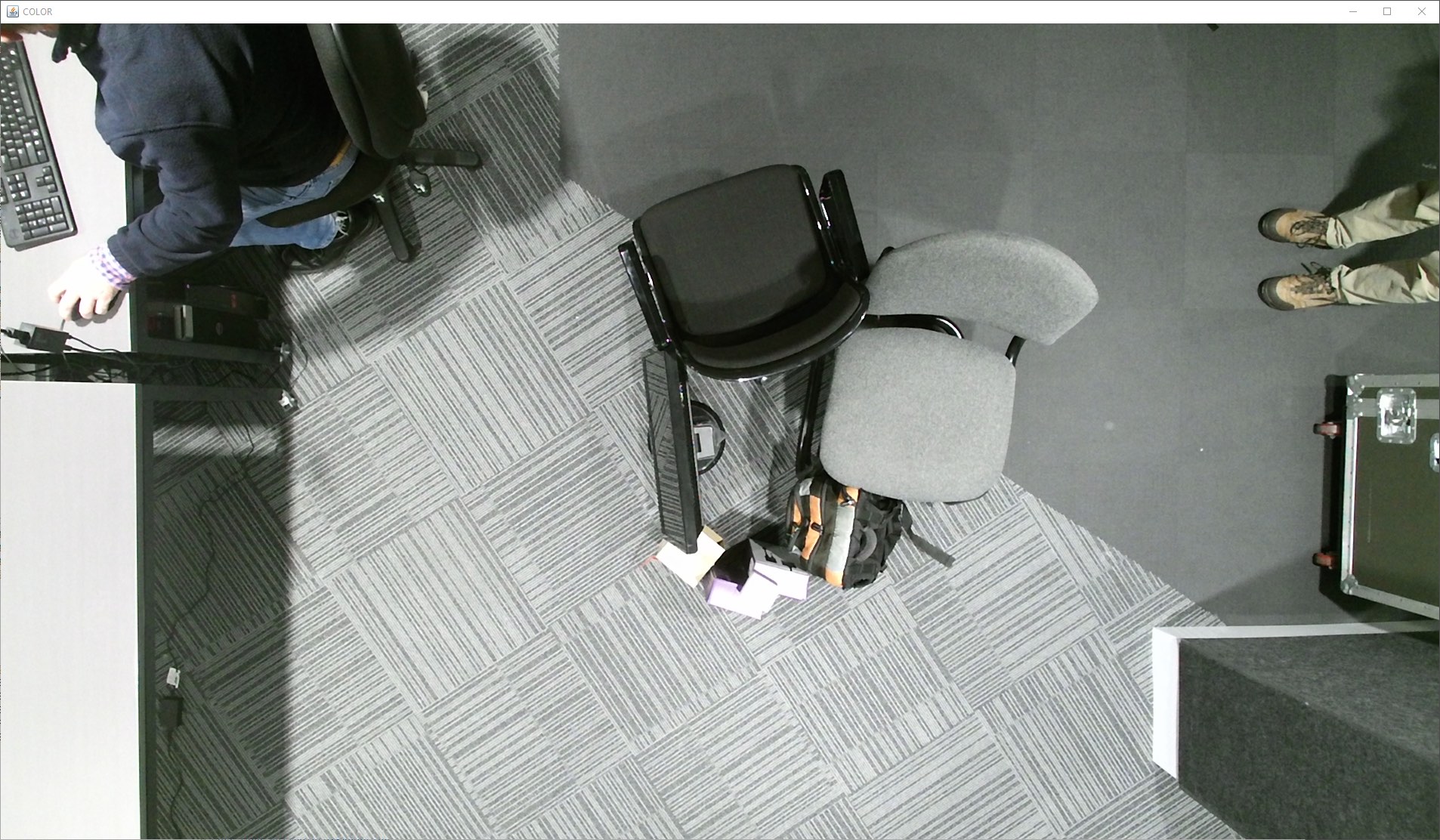}
\includegraphics[width=.3\textwidth]{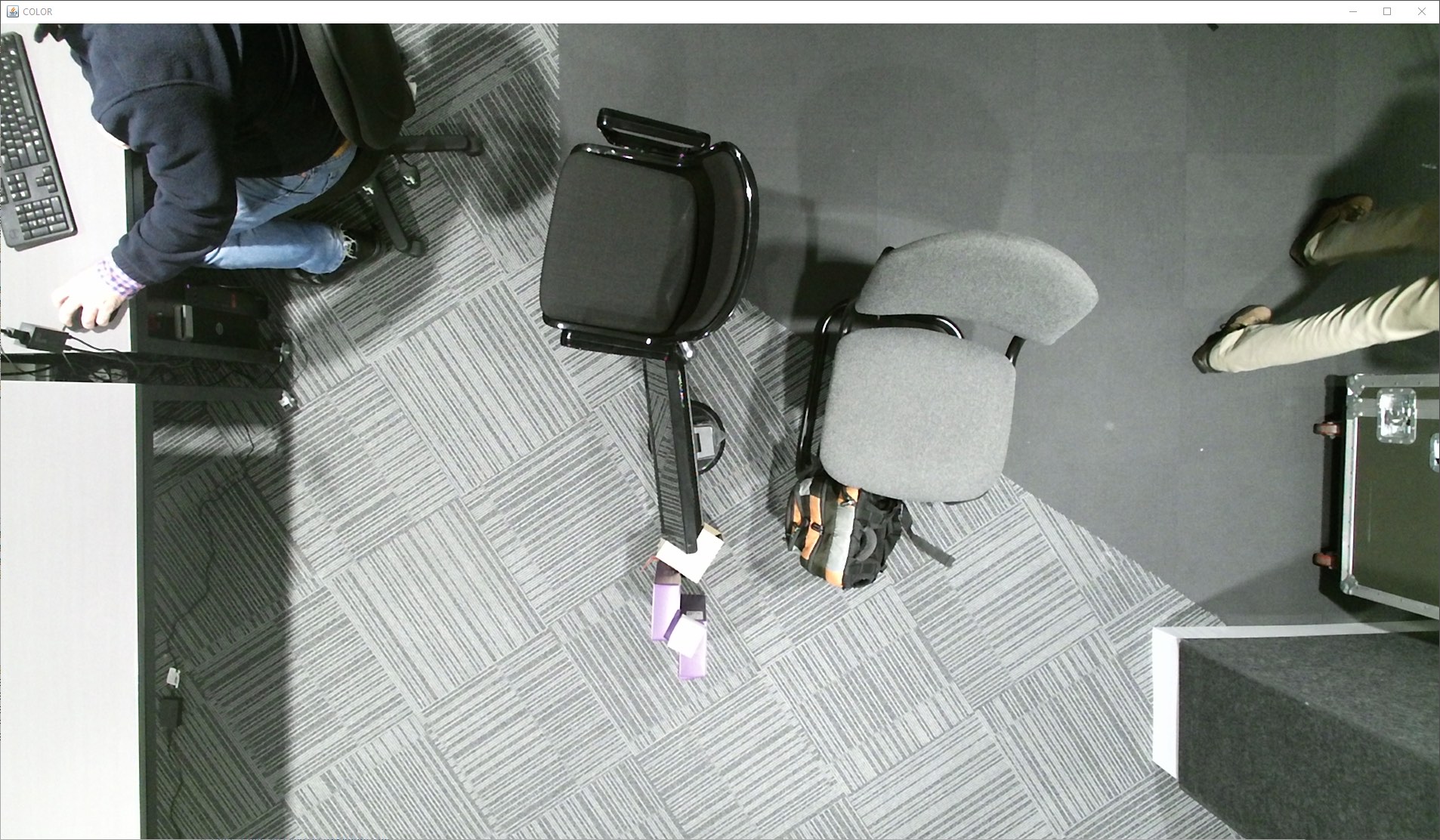}
\includegraphics[width=.3\textwidth]{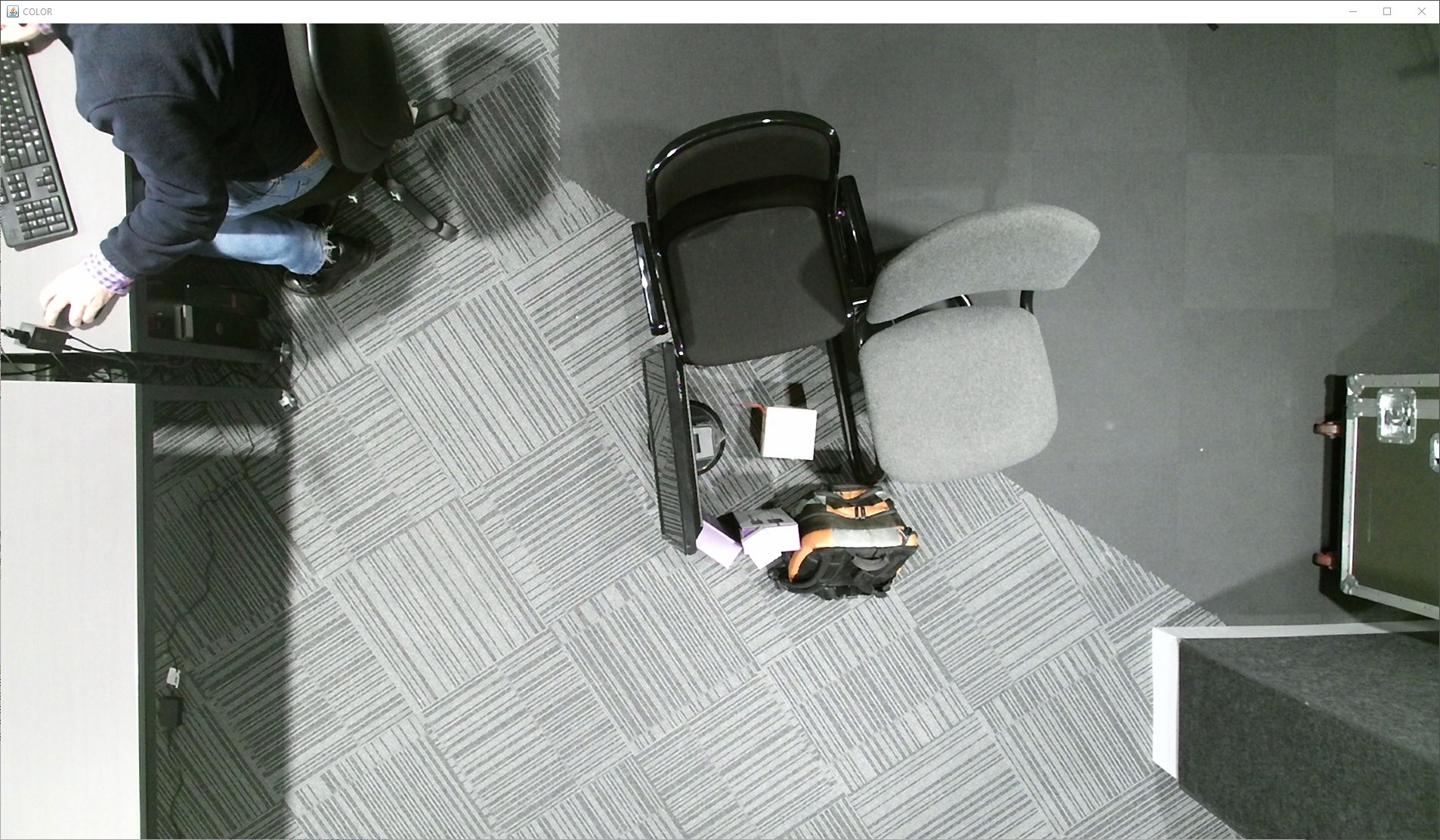}
\includegraphics[width=.3\textwidth]{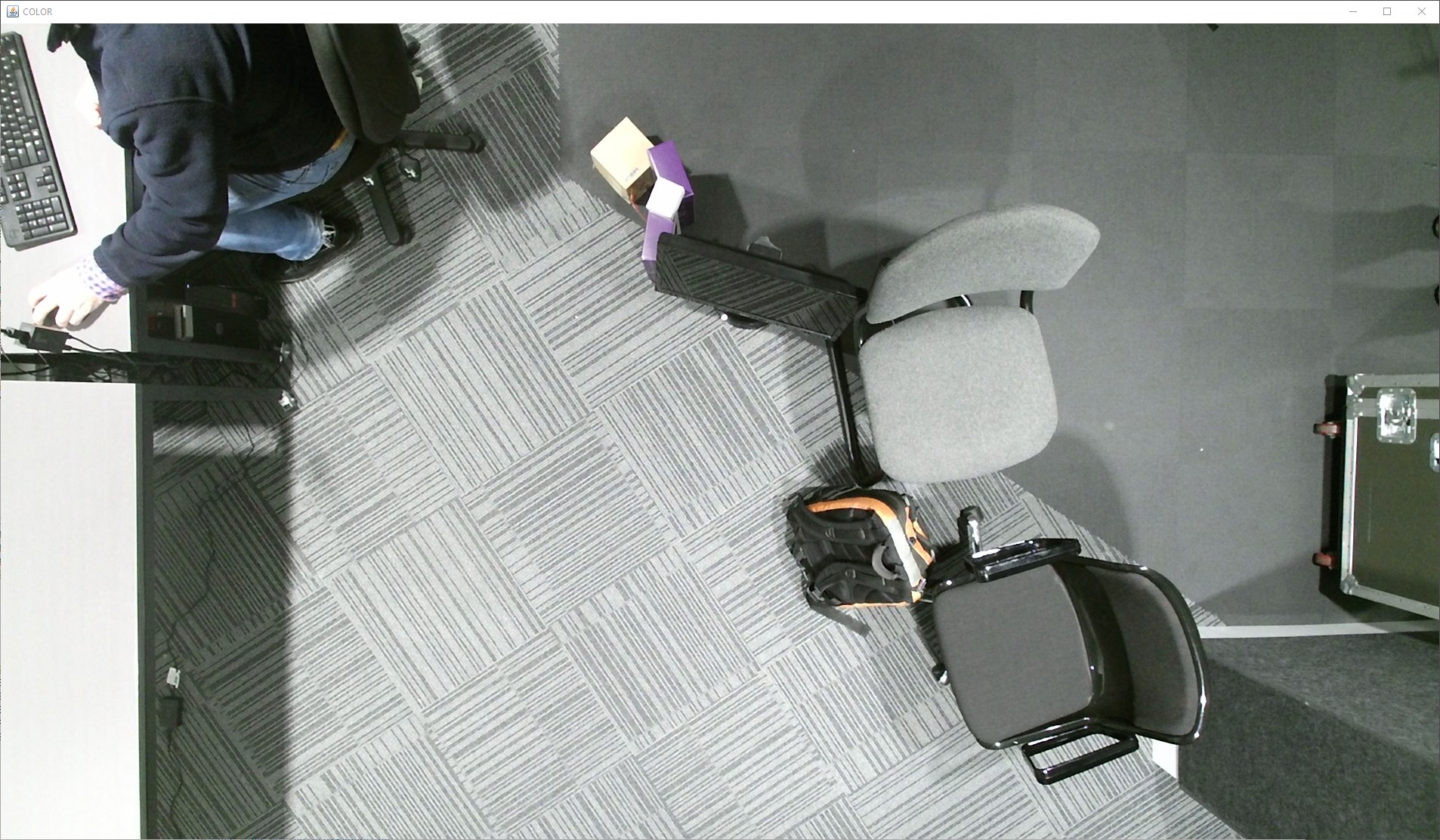}
\includegraphics[width=.3\textwidth]{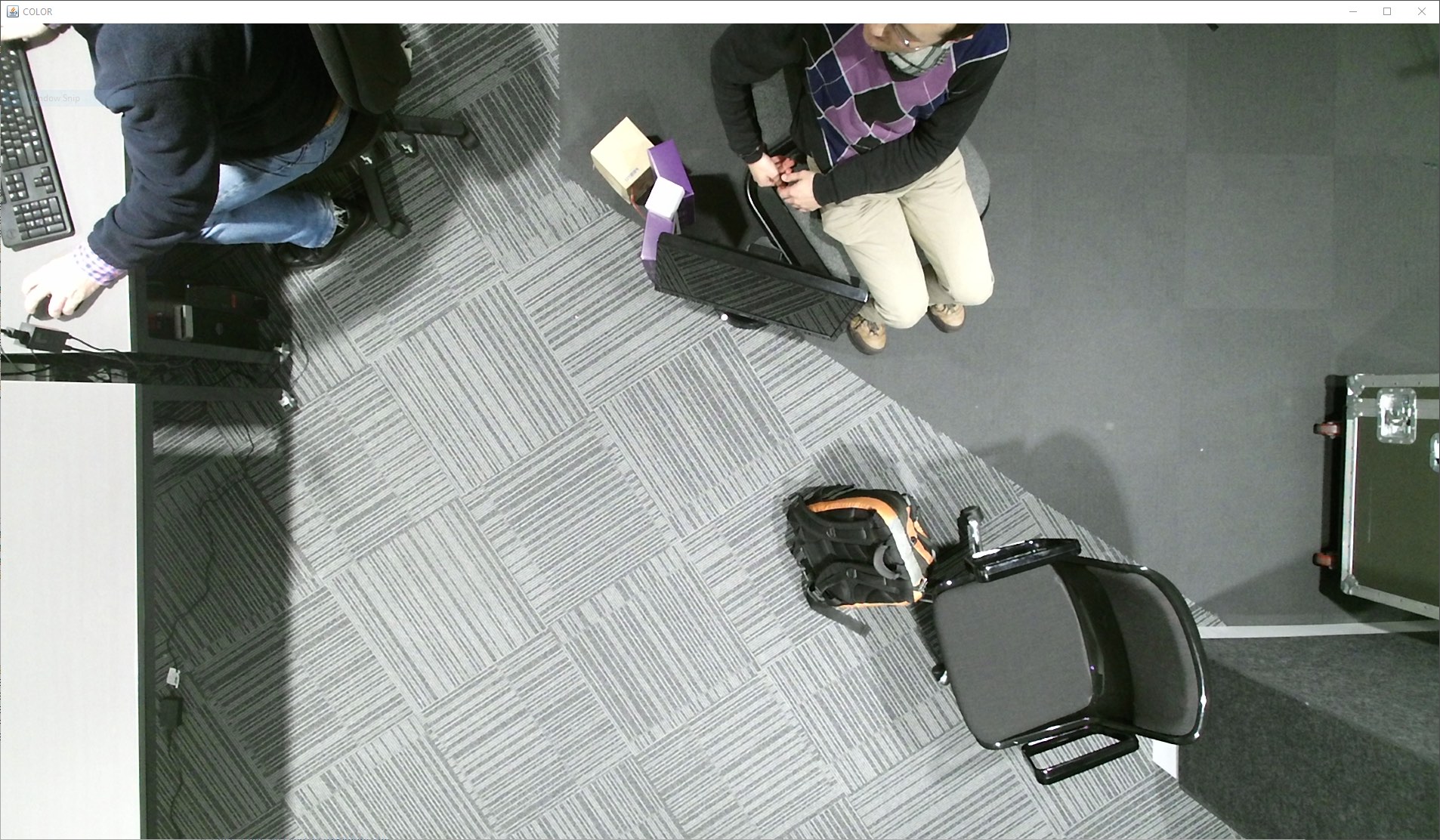}
\caption{Color data for obstacles 1 through 15 from left to right then top down}
\label{fig:obstaclesImages}
\end{figure*}

\begin{figure*}
\centering
\includegraphics[width=.3\textwidth]{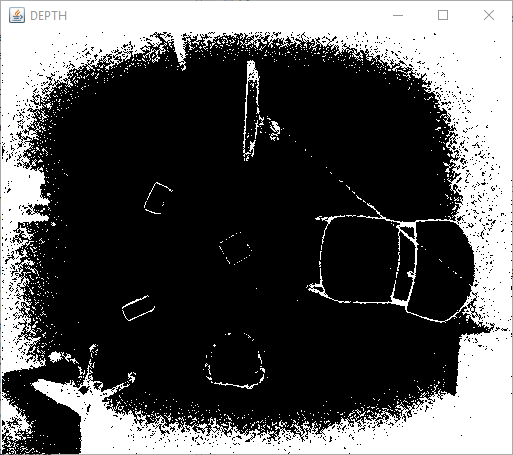}
\includegraphics[width=.3\textwidth]{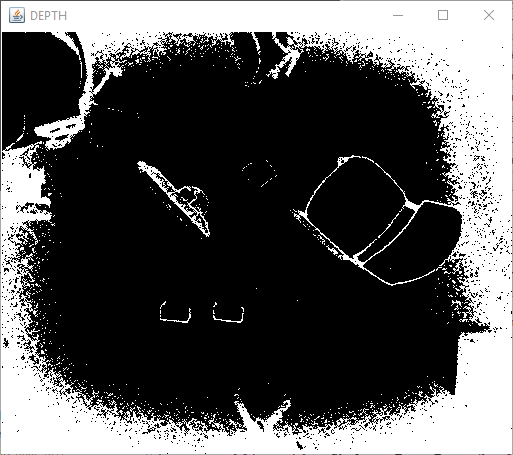}
\includegraphics[width=.3\textwidth]{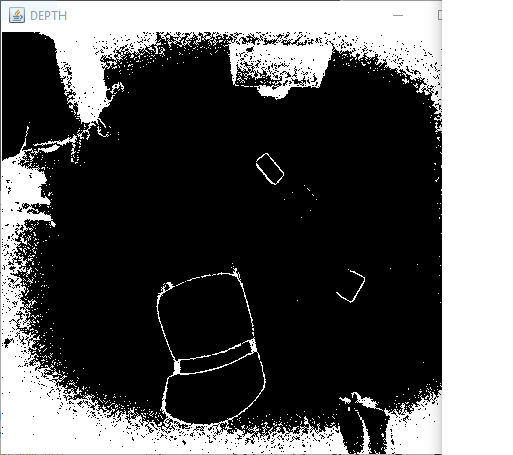}
\includegraphics[width=.3\textwidth]{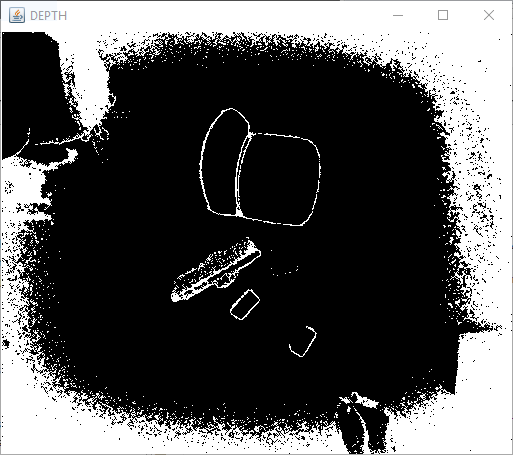}
\includegraphics[width=.3\textwidth]{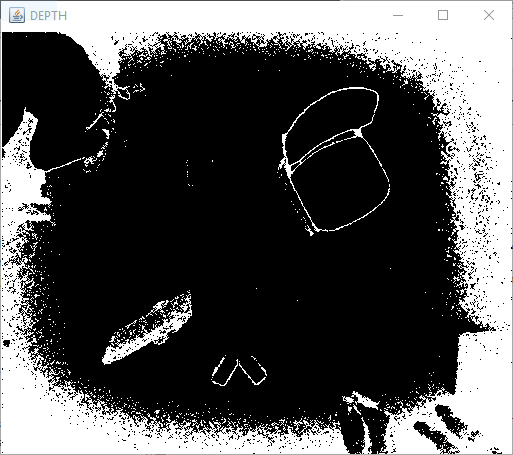}
\includegraphics[width=.3\textwidth]{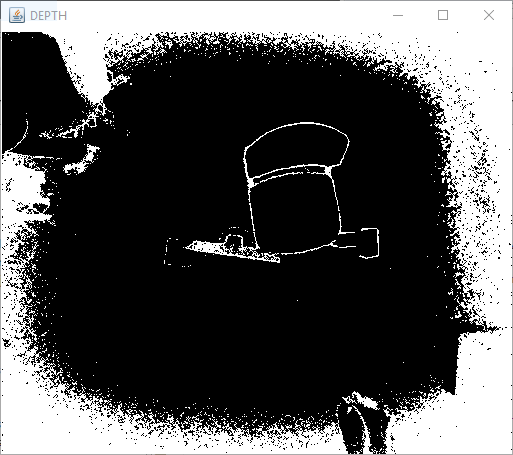}
\includegraphics[width=.3\textwidth]{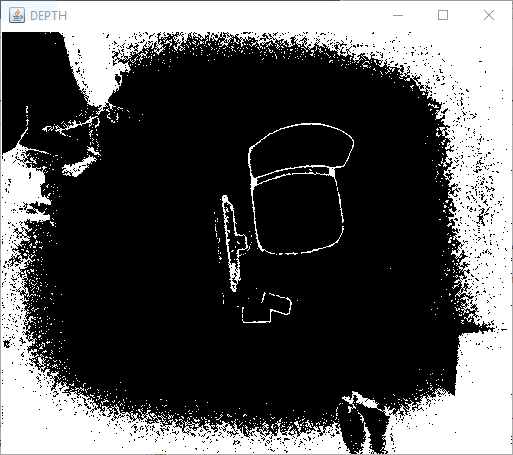}
\includegraphics[width=.3\textwidth]{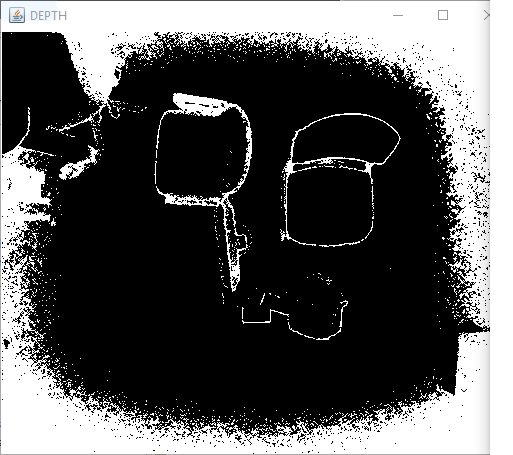}
\includegraphics[width=.3\textwidth]{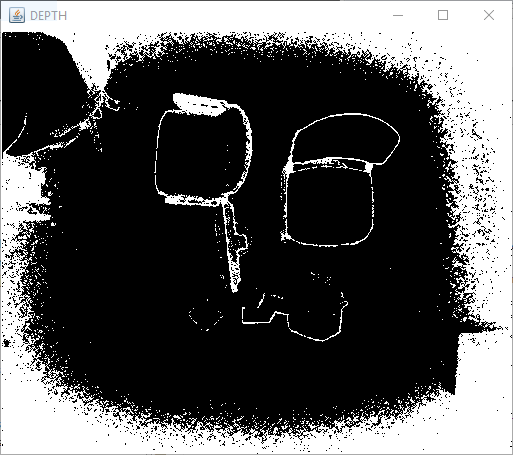}
\includegraphics[width=.3\textwidth]{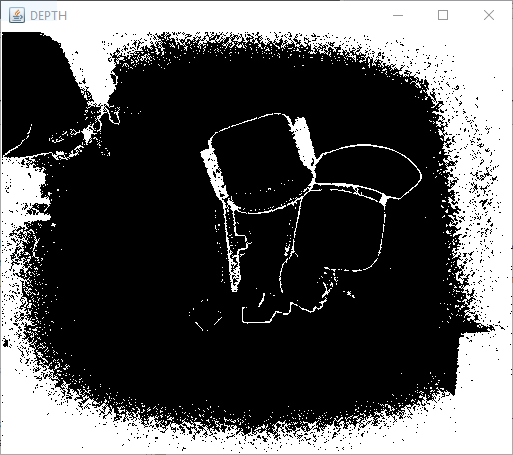}
\includegraphics[width=.3\textwidth]{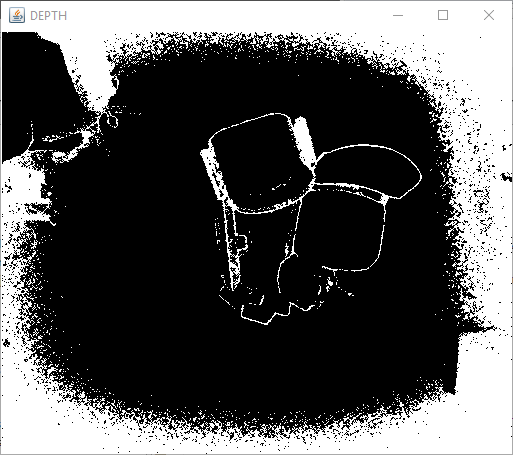}
\includegraphics[width=.3\textwidth]{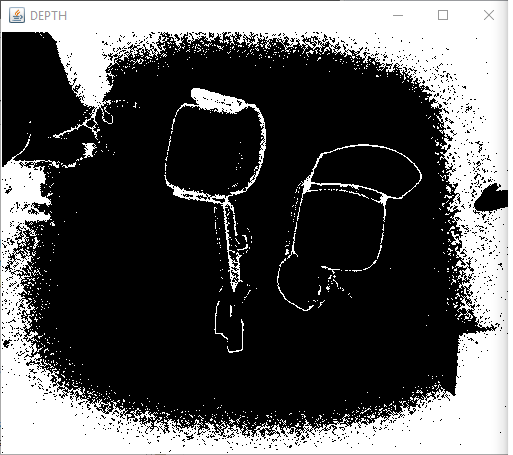}
\includegraphics[width=.3\textwidth]{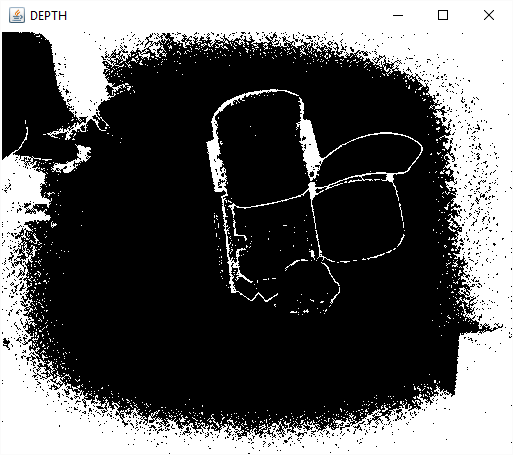}
\includegraphics[width=.3\textwidth]{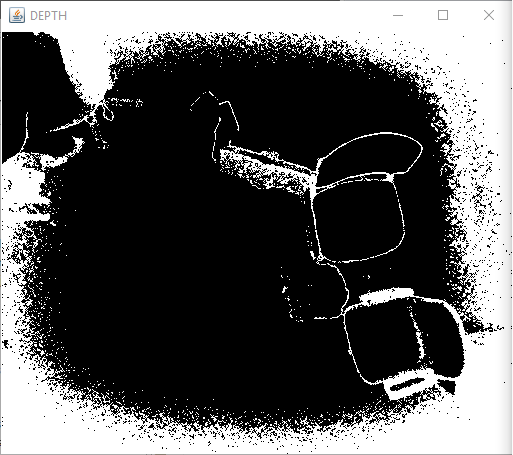}
\includegraphics[width=.3\textwidth]{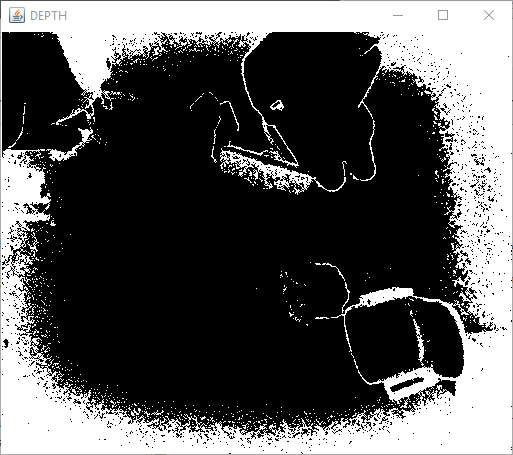}
\caption{Depth data for obstacles 1 through 15 from left to right then top down}
\label{fig:obstaclesImages2}
\end{figure*}

In addition to the data set we have recorded the associated photo(color data) and skeletal (depth) images for each obstacle data set.
These are located on the obstaclesNN directories under the data directory in the BeSpaceD repository.
Figure~\ref{fig:ob1ac} and
Figure~\ref{fig:ob1as} show an example color data set (as a regular
photo image) and the corresponding depth data (as a Kinect skeleton image).
This is for the data set names "obstacles1a".
%
%
%------------------------------------------------------------------ Obstacles 1a
%Image
%
%------------------------------------------------------------------ Obstacles 1 - 15
%
% IMAGES
In Figure~\ref{fig:obstaclesImages} we show the color data for the remaining fifteen obstacles data sets.
In Figure~\ref{fig:obstaclesImages2} we show the depth data for the remaining fifteen obstacles data sets.

%------------------------------------------------------------------------------- Factory Automation

\section{Sensor Data for Factory Automation Collection}
\label{sec:festo}

We gathered live sensor and actuator data from a Festo mini factory owned by
RMIT University in Melbourne, Australia as part of the advanced manufacturing
precinct(AMP).

\begin{figure}
\centering
\includegraphics[width=.4\textwidth]{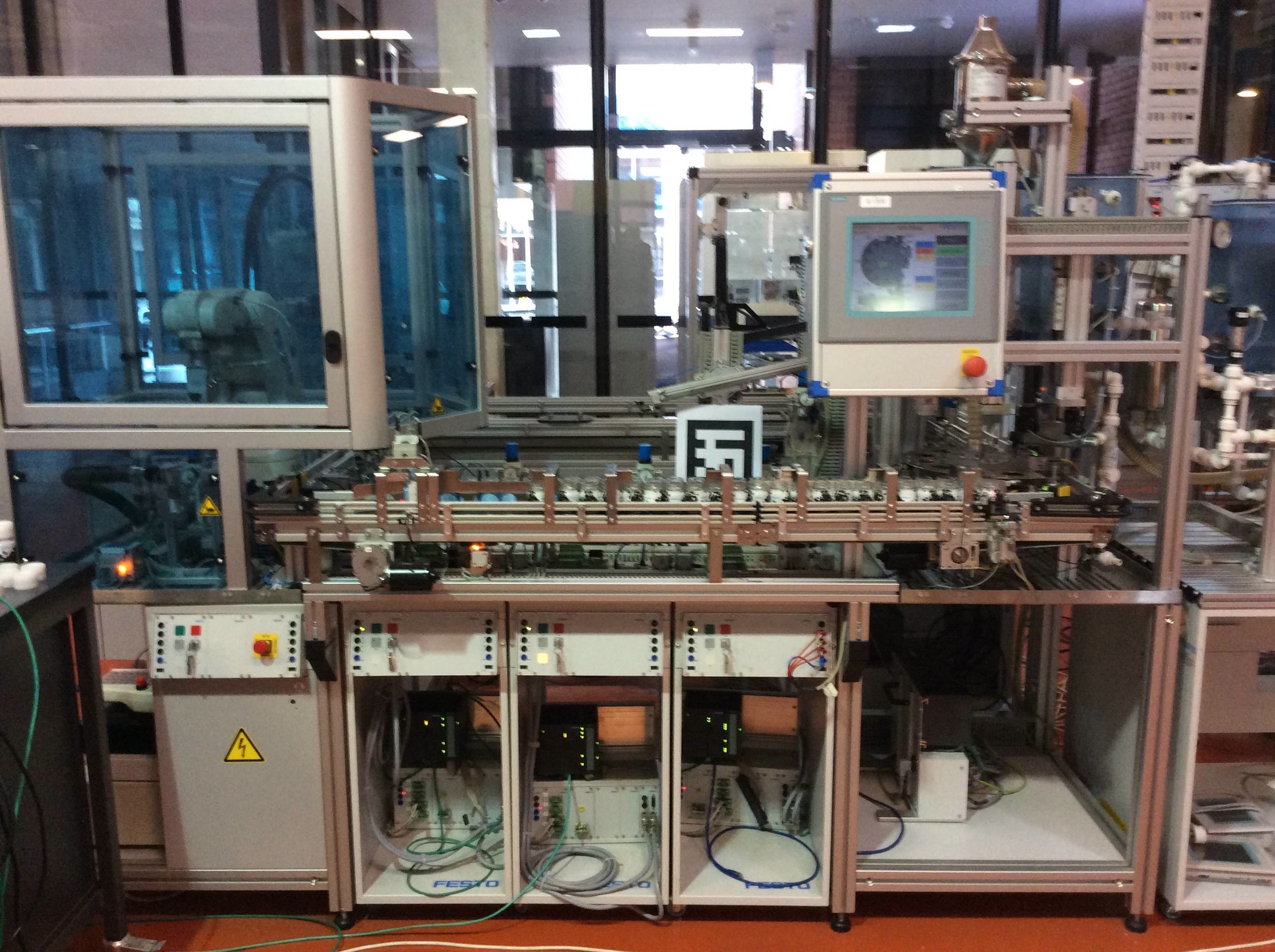}
\includegraphics[width=.55\textwidth]{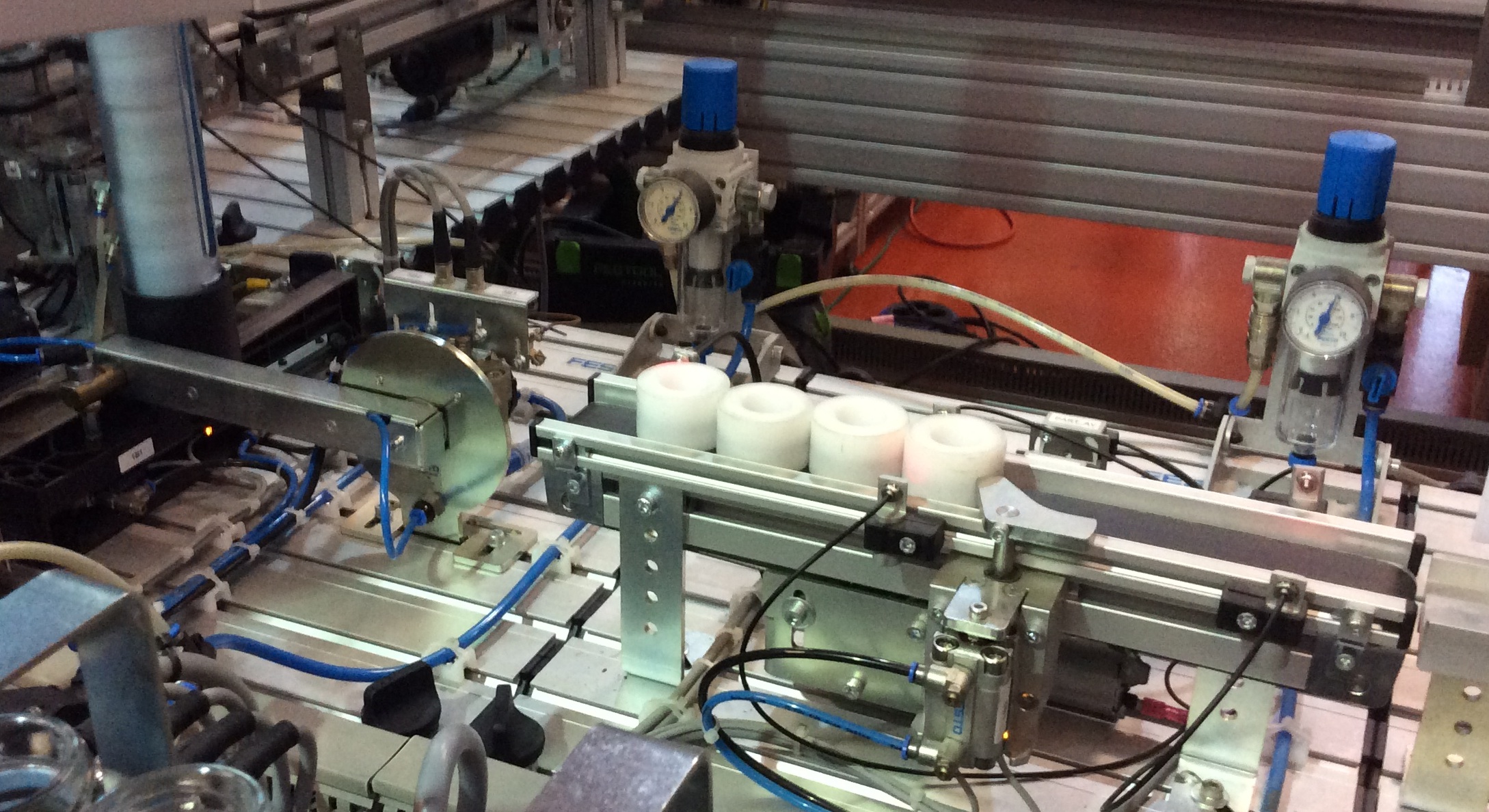}
\caption{The Festo mini factory(left) and a closeup of a station(right)}
\label{fig:festo}
\end{figure}

\subsection{Data description}
For the two data sets below we ran the "FestoRaspberryPiDemonstrator2" program
in the package "BeSpaceDFesto.demonstrator.opcuaclient2channel" in the BeSpaceD repository.
This demo uses a single station of the Festo mini factory only.
The station operates a lever that picks up bottle caps using suction and drops them on a conveyer belt.
There is a vertical tube that takes a ten caps but one or two more can be stacked on top, but this is not recommended.
We did this over stacking a few times in the scenario1 data set.
The station has six sensors and five actuators:

\begin{itemize}
\item Actuators

\begin{itemize}
\item stackEjectorExtendSol : The cap ejector is a horizontally moving part that pushes the single cap at the bottom of the tube out.
\item vacuumGripperSol : Starts/stops the suction in the vacuum gripper which is located at the tip of the lever.
\item ejectionAirPulseSol : Opens the vacuum on the gripper and pushes air through to release the cap.
\item loaderPickupSol : Moves the lever to the cap pickup position.
\item loaderDropoffSol : Moves the level to the cap drop off position (over the conveyer belt). 
\end{itemize}
%KF: The meanings have been confirmed by Yvette

\item Sensors 

\begin{itemize}
\item stackEjectorExtendedLS : Detects when the  cap ejector is in the extended position (i.e. fully pushed the cap out).
\item stackEjectorRetractedLS : Detects when the cap ejector is in the retracted position (i.e. allows all the higher caps to fall down by one cap height).
\item workpieceGrippedSensor : Detects when a cap is fully gripped by suction (i.e. no more air can be extracted).
\item loaderPickupLS : Detects when the lever is in the pickup position (above where the caps exit the tube).
\item loaderDropoffLS : Detects when the lever is in the drop off position (above the conveyer belt).
\item stackEmptySensor : Detects when there are no more caps in the tube.
\end{itemize}
\end{itemize}

The identifiers above correspond to the ComponentState values in the premise of the IMPLIES.
The values of the events are as follows.
For actuators a value of 100 corresponds to a low voltage level and means the actuator is active.
A value of 80 corresponds to a high voltage level and means the actuator is inactive.
For sensors a value of 5.0 is a low signal and a value of 5.5 is a high signal.
The meanings of the high and low signals for each sensor is context dependent.
Generally light sensors are either blocked by a bottle or not.

\subsection{Data structure}
This data describes sensors and actuators changing state in the Festo mini factory during the course of running a demonstrator program that controls the mini factory.
Each sensor data event is represented as a single IMPLIES invariant.
The premise is the conjunction of both time point and the signal identifier. The associated value is the conclusion. For example:
{\small
\begin{verbatim}
    IMPLIES(
        AND(
            TimePoint(Wed Jul 27 09:11:28 UTC 2016),
            Component(stackEjectorExtendSol)
            ),
        ComponentState(5.0)
        )
\end{verbatim}
}
The state value is represented as a ComponentState invariant with floating point values.
The entire event stream for the duration of the capture is represented as a BIGAND of all these implications.
Below is an example of this data as it is constructed in BeSpaceD invariants:

{\footnotesize
\begin{verbatim}
BIGAND(List(
    IMPLIES(AND(TimePoint(Wed Jul 27 09:11:28 UTC 2016),
        Component(stackEjectorExtendSol)),ComponentState(5.0)),
    IMPLIES(AND(TimePoint(Wed Jul 27 09:11:28 UTC 2016),
        Component(stackEjectorExtendedLS)),ComponentState(100.0)),
    IMPLIES(AND(TimePoint(Wed Jul 27 09:11:29 UTC 2016),
        Component(stackEmptySensor)),ComponentState(80.0)),
    IMPLIES(AND(TimePoint(Wed Jul 27 09:11:29 UTC 2016),
        Component(stackEmptySensor)),ComponentState(100.0)),
    IMPLIES(AND(TimePoint(Wed Jul 27 09:11:29 UTC 2016),
        Component(stackEjectorRetractedLS)),ComponentState(80.0)),
    IMPLIES(AND(TimePoint(Wed Jul 27 09:11:29 UTC 2016),
        Component(stackEjectorExtendSol)),ComponentState(5.5)),
    IMPLIES(AND(TimePoint(Wed Jul 27 09:11:29 UTC 2016),
        Component(stackEjectorRetractedLS)),ComponentState(100.0)),
    IMPLIES(AND(TimePoint(Wed Jul 27 09:11:29 UTC 2016),
        Component(stackEmptySensor)),ComponentState(80.0)),
    IMPLIES(AND(TimePoint(Wed Jul 27 09:11:30 UTC 2016),
        Component(stackEjectorExtendedLS)),ComponentState(80.0)),
    IMPLIES(AND(TimePoint(Wed Jul 27 09:11:30 UTC 2016),
        Component(loaderPickupSol)),ComponentState(5.0)),
    IMPLIES(AND(TimePoint(Wed Jul 27 09:11:30 UTC 2016),
        Component(stackEmptySensor)),ComponentState(100.0)),
    IMPLIES(AND(TimePoint(Wed Jul 27 09:11:30 UTC 2016),
        Component(loaderPickupLS)),ComponentState(80.0)),
    IMPLIES(AND(TimePoint(Wed Jul 27 09:11:30 UTC 2016),
        Component(loaderPickupSol)),ComponentState(5.5)),
    IMPLIES(AND(TimePoint(Wed Jul 27 09:11:30 UTC 2016),
        Component(vacuumGripperSol)),ComponentState(5.0)),
    IMPLIES(AND(TimePoint(Wed Jul 27 09:11:31 UTC 2016),
        Component(workpieceGrippedSensor)),ComponentState(80.0)),
    IMPLIES(AND(TimePoint(Wed Jul 27 09:11:31 UTC 2016),
        Component(workpieceGrippedSensor)),ComponentState(80.0)),
    IMPLIES(AND(TimePoint(Wed Jul 27 09:11:31 UTC 2016),
        Component(workpieceGrippedSensor)),ComponentState(100.0)),
    IMPLIES(AND(TimePoint(Wed Jul 27 09:11:31 UTC 2016),
        Component(loaderDropoffSol)),ComponentState(5.0)),
    IMPLIES(AND(TimePoint(Wed Jul 27 09:11:31 UTC 2016),
        Component(loaderPickupLS)),ComponentState(100.0)),
    IMPLIES(AND(TimePoint(Wed Jul 27 09:11:33 UTC 2016),
        Component(loaderDropoffLS)),ComponentState(80.0)),
    IMPLIES(AND(TimePoint(Wed Jul 27 09:11:33 UTC 2016),
        Component(loaderDropoffSol)),ComponentState(5.5)),
    IMPLIES(AND(TimePoint(Wed Jul 27 09:11:33 UTC 2016),
        Component(vacuumGripperSol)),ComponentState(5.5)),
    IMPLIES(AND(TimePoint(Wed Jul 27 09:11:33 UTC 2016),
        Component(ejectionAirPulseSol)),ComponentState(5.0)),
    IMPLIES(AND(TimePoint(Wed Jul 27 09:11:33 UTC 2016),
        Component(workpieceGrippedSensor)),ComponentState(100.0)),
    IMPLIES(AND(TimePoint(Wed Jul 27 09:11:34 UTC 2016),
        Component(loaderPickupSol)),ComponentState(5.0)),
    IMPLIES(AND(TimePoint(Wed Jul 27 09:11:34 UTC 2016),
        Component(ejectionAirPulseSol)),ComponentState(5.5)),
    IMPLIES(AND(TimePoint(Wed Jul 27 09:11:34 UTC 2016),
        Component(loaderDropoffLS)),ComponentState(100.0)),
    IMPLIES(AND(TimePoint(Wed Jul 27 09:11:35 UTC 2016),
        Component(stackEjectorExtendSol)),ComponentState(5.0)),
    ...
))
\end{verbatim}
}

\noindent This is just an excerpt of the first twenty eight events of the actual
data set which contains a total of 4761 events.
The subset of events above roughly covers every actuator and sensor in one complete cycle of the system.

\subsection{Data set access}
In order to access the train occupancy data set one can call the following:
\begin{verbatim}
      Robotics.Festo.MiniFactory.station1.scenario1()
      Robotics.Festo.MiniFactory.station1.capsBlocking()
\end{verbatim}
In addition to the BeSpaceD data sets above we also have some associated data
recorded in the "data" directory of the BeSpaceD repository:
\begin{verbatim}
    aicause.festo.station1.Scenario1.20mins      // The archived data set
    aicause.festo.station1.Scenario1.20mins.README
    aicause.festo.station1.Scenario1.20mins.log
    aicause.festo.station1.Scenario1.20mins.txt
    aicause.festo.station1.small.2capsBlocking
\end{verbatim}
The base files are the data set archives.
The ".README" file describes what was done manually for the Scenario 1 data set.
The ".log" file contains the output of the Festo demonstrator including state transition
information and each signal transformed into a BeSpaceD invariant.
The ".txt" file is a readable version of the archived data set.

%------------------------------------------------------------------------------- Train Occupancy

\section{Train Occupancy Collection}
\label{sec:train}
We gathered train occupancy data from a separate research project
involving Lego Train sets in Norway controlled by reactive blocks \cite{reactive}  and remote monitoring in
Australia (see \cite{trains1} and \cite{trains2}). 

\subsection{Data description}
In this project there are two parts to the data:
The layout of the track network and the position of the train on the track.
The track network is make up of individual lego track pieces.
This is represented symbolically as track segments, each with a unique integer.
The static track layout is not part of the data we collected for BeSpaceD.
The train occupies ten track segments and therefore can be represented as a set of these track segments.

\subsection{Data structure}
This data describes the spatial location of the train symbolically.
Each time point and the associated occupancy is represented as a single IMPLIES invariant.
{\small
\begin{verbatim}
    IMPLIES(TimePoint(1429188806320), BIGAND... )
\end{verbatim}
}
The time point is the premise and the occupancy is the conclusion.
The occupancy data is represented as a conjunction (BIGAND) of ten OccupyNode invariants.
{\small
\begin{verbatim}
    BIGAND(List(
        OccupyNode(664), OccupyNode(665), OccupyNode(666), OccupyNode(667),
        OccupyNode(668), OccupyNode(669), OccupyNode(670), OccupyNode(671),
        OccupyNode(672), OccupyNode(1)
        ))
\end{verbatim}
}
Normally these OccupyNodes will be adjacent in the semantic description of the train track network.
Therefore the ten OccupyNodes represent the location of one train on the track.
The entire train occupancy invariant over all time points is represented as a BIGAND of all the implications.
Below is an example of this data as it is constructed in BeSpaceD invariants:

{\small
\begin{verbatim}
BIGAND(List(
IMPLIES(TimePoint(1429188806320),BIGAND(List(OccupyNode(664), OccupyNode(665), 
        OccupyNode(666), OccupyNode(667), OccupyNode(668), OccupyNode(669), 
        OccupyNode(670), OccupyNode(671), OccupyNode(672), OccupyNode(1)))),
IMPLIES(TimePoint(1429188806417),BIGAND(List(OccupyNode(665), OccupyNode(666), 
        OccupyNode(667), OccupyNode(668), OccupyNode(669), OccupyNode(670), 
        OccupyNode(671), OccupyNode(672), OccupyNode(1), OccupyNode(2)))),
IMPLIES(TimePoint(1429188806527),BIGAND(List(OccupyNode(666), OccupyNode(667), 
        OccupyNode(668), OccupyNode(669), OccupyNode(670), OccupyNode(671), 
        OccupyNode(672), OccupyNode(1), OccupyNode(2), OccupyNode(3)))),
IMPLIES(TimePoint(1429188806608),BIGAND(List(OccupyNode(667), OccupyNode(668), 
        OccupyNode(669), OccupyNode(670), OccupyNode(671), OccupyNode(672), 
        OccupyNode(1), OccupyNode(2), OccupyNode(3), OccupyNode(4)))),
IMPLIES(TimePoint(1429188806715),BIGAND(List(OccupyNode(668), OccupyNode(669), 
        OccupyNode(670), OccupyNode(671), OccupyNode(672), OccupyNode(1), 
        OccupyNode(2), OccupyNode(3), OccupyNode(4), OccupyNode(5)))),
IMPLIES(TimePoint(1429188806799),BIGAND(List(OccupyNode(669), OccupyNode(670), 
        OccupyNode(671), OccupyNode(672), OccupyNode(1), OccupyNode(2), 
        OccupyNode(3), OccupyNode(4), OccupyNode(5), OccupyNode(6)))),
IMPLIES(TimePoint(1429188806893),BIGAND(List(OccupyNode(670), OccupyNode(671), 
        OccupyNode(672), OccupyNode(1), OccupyNode(2), OccupyNode(3), 
        OccupyNode(4), OccupyNode(5), OccupyNode(6), OccupyNode(7)))),
IMPLIES(TimePoint(1429188807001),BIGAND(List(OccupyNode(671), OccupyNode(672), 
        OccupyNode(1), OccupyNode(2), OccupyNode(3), OccupyNode(4), 
        OccupyNode(5), OccupyNode(6), OccupyNode(7), OccupyNode(8)))),
IMPLIES(TimePoint(1429188807096),BIGAND(List(OccupyNode(672), OccupyNode(1), 
        OccupyNode(2), OccupyNode(3), OccupyNode(4), OccupyNode(5), 
        OccupyNode(6), OccupyNode(7), OccupyNode(8), OccupyNode(9)))),
IMPLIES(TimePoint(1429188807191),BIGAND(List(OccupyNode(1), OccupyNode(2), 
        OccupyNode(3), OccupyNode(4), OccupyNode(5), OccupyNode(6), 
        OccupyNode(7), OccupyNode(8), OccupyNode(9), OccupyNode(10)))),
...
)
\end{verbatim}
}
This is just an excerpt of the first ten time points of the actual data which contains a total of 9601 time points.

\subsection{Data set access}
In order to access the train occupancy data set one can call the following:
\begin{verbatim}
    Robotics.Lego.Trains.experiment1()
\end{verbatim}

%------------------------------------------------------------------------------- Weather Data

\section{Weather Data Collection}
\label{sec:wet}

We gathered Ultra Violet (UV) index data from a separate research
project called SmartSpace involving ABB's research centre in India
(INCRC) and AICAUSE in Melbourne, Australia. The data sets have been
used in the SmartSpace project (see \cite{smartspace} and \cite{smartspace3d}).

\subsection{Data description}
The standard way to represent the UV Index is a value from 0 and 10 inclusive
with up to two decimal places.
We used the ARPANSA website to collect UV data\footnote{\url{www.arpansa.gov.au/uvindex/realtime/xml}}.
In ARPANSA, indexes only have one decimal place and are represented as an integer
where the index is multiplied by 100. 
In BeSpaceD we retain the same integer values
so a value of 770 represents a UV index of 7.7.

\subsection{Data structure}
Each time point and the associated UV data is represented as a single IMPLIES invariant.
The time point is the premise and the UV data is the conclusion.
The UV data data is represented as a conjunction (BIGAND) of four ComponentState invariants.
These are effectively the values of the UV data structure.
The value of interest is the UV Index identified with the premise of Owner("Index").
Other data such as the name and ID are also recorded with the index.
The invariant for the entire UV data for one city over all time points is represented as a BIGAND of all the implications.

Below is an example of this data as it is constructed in BeSpaceD invariants.
This example data contains the UV Index for Melbourne, Australia at many times during the day down to a one second resolution:

\begin{verbatim}
BIGAND(List(
IMPLIES(TimePoint(1st December 201511:04AM),
    BIGAND(List(
        IMPLIES(Owner(ID),ComponentState(melbourne)),
        IMPLIES(Owner(Index),ComponentState(770)),
        IMPLIES(Owner(Name),ComponentState(mel))))),

IMPLIES(TimePoint(8th December 201511:05AM),
    BIGAND(List(
        IMPLIES(Owner(ID),ComponentState(melbourne)),
        IMPLIES(Owner(Index),ComponentState(460)),
        IMPLIES(Owner(Name),ComponentState(mel))))),

IMPLIES(TimePoint(8th December 201511:06AM),
    BIGAND(List(
        IMPLIES(Owner(ID),ComponentState(melbourne)),
        IMPLIES(Owner(Index),ComponentState(480)),
        IMPLIES(Owner(Name),ComponentState(mel))))),

IMPLIES(TimePoint(3rd December 201511:07AM),
    BIGAND(List(
        IMPLIES(Owner(ID),ComponentState(melbourne)),
        IMPLIES(Owner(Index),ComponentState(780)),
        IMPLIES(Owner(Name),ComponentState(mel))))),

IMPLIES(TimePoint(3rd December 201511:08AM),
    BIGAND(List(
        IMPLIES(Owner(ID),ComponentState(melbourne)),
        IMPLIES(Owner(Index),ComponentState(790)),
        IMPLIES(Owner(Name),ComponentState(mel))))),

IMPLIES(TimePoint(3rd December 201511:09AM),
    BIGAND(List(
        IMPLIES(Owner(ID),ComponentState(melbourne)),
        IMPLIES(Owner(Index),ComponentState(790)),
        IMPLIES(Owner(Name),ComponentState(mel))))),

IMPLIES(TimePoint(3rd December 201511:10AM),
    BIGAND(List(
        IMPLIES(Owner(ID),ComponentState(melbourne)),
        IMPLIES(Owner(Index),ComponentState(790)),
        IMPLIES(Owner(Name),ComponentState(mel))))),

IMPLIES(TimePoint(3rd December 201511:11AM),
    BIGAND(List(
        IMPLIES(Owner(ID),ComponentState(melbourne)),
        IMPLIES(Owner(Index),ComponentState(800)),
        IMPLIES(Owner(Name),ComponentState(mel))))),
...
)
\end{verbatim}

This is just an excerpt of the first eight time points of the actual data which contains a total of 439 time points.

\subsection{Data set access}
Currently we only have one weather data set but plan to add more including precipitation radar.
In order to access the UV weather data set one can call the following:

\begin{verbatim}
    Weather.SmartSpace.Melbourne.uvIndex_Dec_28_2015()
\end{verbatim}

%------------------------------------------------------------------------------- Conclusion

\section{Conclusion}
\label{sec:concl}

In this paper we have described the processes of accessing and producing
BeSpaceD data sets for experimental purposes.
These data sets are organised into collections with a common data model
and further organised into and ontology making them easy to access from Scala.
In addition, we are providing to the community a number of varied data sets that
we and our collaborators found useful.
We describe the data model for each collection and give technical details about how to access them.
In some instances we provide images or log files associated with the data as a reference.
We hope this data will be beneficial to the community and the API we developed
will encourage more collaborators to contribute more and varied data sets
with interesting data models in BeSpaceD.

%------------------------------------------------------------------------------- Acknowledgement

\subsection*{Acknowledgement}
The authors would like to thank Simon Hordvik, Kristoffer {\O}seth,
and Peter Herrmann from NTNU for the provisioning of the train data.
We also thank Yvette Wouters and Lasith Fernando from RMIT for
creating the Festo demonstrator software and hardware respectively.
We'd like to thank Alexander Yeap, Vasileios Dimitrakopoulos,
Nicholas Powell and Shane Basnayake from RMIT for creating the
basis of the Kinect scanning software and provisioning the bottle scan.
We also thank Zoran Savic for his support with the Festo mini factory and Ian Peake for his support in the VXLab.

%------------------------------------------------------------------------------- Bibliography

 \bibliographystyle{eptcs}

\end{document}